\documentclass[twocolumn,prb,reprint,amsmath,amssymb,aps,showpacs]{revtex4-1}

\usepackage{graphicx}
\usepackage{dcolumn}
\usepackage{bm}
\usepackage{graphics}
\usepackage{epsfig}
\usepackage{epstopdf}
\begin{document}

\title{Bose-Einstein Condensation of two-dimensional polaritons in microcavity   under the influence of the Landau quantization and Rashba spin-orbit coupling}

\author{S.A. Moskalenko,$^{1}$ E.V. Dumanov,$^{1,2}$ I.V. Podlesny,$^{1}$
M.A. Liberman,$^{3,4}$ B.V. Novikov,$^{5}$ S.S. Rusu,$^{1,2}$ V.M. Bajireanu$^{1}$}
\affiliation{$^{1}$Institute of Applied Physics of the Academy of Sciences of Moldova, Academic Str. 5, Chisinau, MD2028, Republic of Moldova\\
$^{2}$Technical University of Moldova, bd. Stefan cel Mare 168, MD2004, Chisinau, Republic of Moldova\\
$^{3}$Nordita, KTH Royal Institute of Technology and Stockholm University, Roslagstullsbacken 23, 10691 Stockholm, Sweden\\
$^{4}$Moscow Institute of Physics and Technology, Inststitutskii per. 9, Dolgoprudnyi, Moskovsk. obl.,141700, Russia\\
$^{5}$Department of Solid State Physics, Institute of Physics, St.Petersburg State University, 1, Ulyanovskaya str., Petrodvorets, 198504 St. Petersburg, Russia}

\date{\today}

\begin{abstract}
The Bose-Einstein condensation (BEC) of the two-dimensional (2D) magnetoexciton-polaritons in microcavity, when the Landau quantization of the electron and hole states accompanied by the Rashba spin-orbit coupling plays the main role, were  investigated. The Landau quantization levels of the 2D heavy holes with nonparabolic dispersion law and third order chirality terms both induced by the external electric field perpendicular to the semiconductor quantum well as the strong magnetic field B gives rise to the nonmonotous dependence on B of the magnetoexciton energy levels and of the polariton energy branches. The Hamiltonian describing the Coulomb electron - electron and the electron - radiation interactions was expressed in terms of the two-particle integral operators such as the density operators $\hat{\rho }(\vec{Q})$ and $\hat{D}(\vec{Q})$ representing the optical and the acoustical plasmons and the magnetoexciton creation and annihilation operators $\Psi _{ex}^{\dagger }({{\vec{k}}_{||}}),\Psi _{ex}^{{}}({{\vec{k}}_{||}})$ with in - plane wave vectors ${{\vec{k}}_{||}}$ and $\vec{Q}$. The polariton creation and annihilation operators $L_{ex}^{\dagger }({{\vec{k}}_{||}}),L_{ex}^{{}}({{\vec{k}}_{||}})$ were introduced using the Hopfield coefficients and neglecting the antiresonant terms because the photon energies exceed the energy of the cavity mode. The BEC of the magnetoexciton - polariton takes place on the lower polariton branch in the point ${{\vec{k}}_{||}}=0$ with the quantized value of the longitudinal component of the light wave vector, as in the point of the cavity mode.
	The unitary coherent transformation of the obtained Hamiltonian leading to the breaking of its gauge symmetry was written as the Glauber - type coherent transformation using the polariton operators $L_{0}^{\dagger },L_{0}^{{}}$ instead of the true Bose operators. It can be represented in a factorized form as a product of two unitary transformations acting separately on the magnetoexciton and photon subsystems. First of them is similar with the Kaldysh-Kozlov-Kapaev unitary transformation whereas the second one is equivalent to the Bogoliubov canonical displacement transformation. It was shown that the first transformation leads not only to the Bogoliubov u-v transformations of the electron and hole single-particle Fermi operators but also to the similar transformation of the two-particle integral operators. It becomes possible due to the extensive N-fold degeneracy of the lowest Landau levels (LLLs) in Landau gauge description, where N is proportional to the layer surface area S. In both cases, the u-v coefficients depend on the LLLs filling factor but in the last case, this dependence is doubled. The breaking of the gauge symmetry gives rise to the new mixed states expressed through the coherent superposition of the algebraic sum of the magnetoexciton creation and of annihilation operators $({{e}^{i\alpha }}\Psi _{ex}^{{}}(-{{\vec{k}}_{||}})+\Psi _{ex}^{\dagger }({{\vec{k}}_{||}}){{e}^{-i\alpha }})$ and of the density operator $\hat{D}({{\vec{k}}_{||}})$ representing the acoustical plasmon. In difference on them the density operator $\hat{\rho }(\vec{Q})$ representing the optical plasmon does not take part in such superpositions.
\end{abstract}

\maketitle
\tableofcontents


\section{Introduction}

The present article is based on the background previous papers and monographs [1-16] as well on the recent contribution [17-26].

In Ref.[17] the Hamiltonian of the electron-radiation interaction in the second quantization representation for the case of 2D coplanar electron-hole (e-h) systems in a strong perpendicular magnetic field was derived. The s-type conduction band electrons with spin projections ${{s}_{z}}=\pm 1/2$ along the magnetic field direction and the heavy holes with total momentum projections ${{j}_{z}}=\pm 3/2$ in the p-type valence band were taken into account. The periodic parts of their Bloch wave functions are similar to $(x\pm iy)$ expressions with the orbital momentum projection ${{M}_{v}}=\pm 1$ on the same selected direction. The envelope parts of the Bloch wave functions have the forms of plane waves in the absence of the magnetic field. In its presence they are completely changed due to the Landau quantization event. In the papers [17-26] the Landau quantization of the 2D electrons and holes is described in the Landau gauge and is characterized by the oscillator-type motion in one in-plane direction giving rise to discrete Landau levels enumerated by the quantum numbers ${{n}_{e}}$ and ${{n}_{h}}$ and by the free translation motion in another in-plane direction perpendicular to previous one. The one-dimensional (1D) plane waves describing this motion are marked by the 1D wave numbers p and q. In Ref.[18] the Landau quantization of the 2D electrons with non-parabolic dispersion law, pseudospin components and chirality terms was investigated. On this base in Ref.[19] the influence of the Rashba spin-orbit coupling (RSOC) on the 2D magnetoexcitons was discussed. The spinor-type wave functions of the conduction and valence electrons in the presence of the RSOC have different numbers of Landau quantization levels for different spin projections. As was  demonstrated in Refs[18, 19, 22], the difference between these numbers is determined by the order of the chirality terms. Their origin is due to the influence of the external electric field applied to the layer parallel to the direction of the magnetic field. In Ref.[19] two lowest Landau levels (LLLs) of the conduction electron and four LLLs for the holes were used to calculate the matrix elements of the Coulomb interaction between the charged carriers as well as the matrix elements of the electron-radiation interaction. On these bases the ionization potentials of the new-magnetoexcitons and the probabilities of the quantum transitions from the ground state of the crystal to the magnetoexciton states were calculated. In the present description the number of the hole and magnetoexciton states will be enlarged and the formation of magnetopolaritons taking into account the RSOC will be described. The more simple variant of the magnetopolariton without taking into account the RSOC was described in Ref.[21] for the case of inter-band quantum transitions and in Ref.[23] for the case of intra-band quantum transitions.

The paper is organized as follows. In the section 2 the results concerning the Landau quantization of the 2D heavy holes as well as of the electrons in the conduction band taking into account the Rashba spin-orbit interaction were remembered. On this base the Hamiltonians describing the electron-radiation interaction and of the Coulomb electron-electron interaction in the presence of the Rashba spin-orbit coupling were deduced in the section 3 and 4 correspondingly. The section 5 is devoted to the description of the magnetoexcitons in the model of a Bose gas. In the section 6 the breaking of the gauge symmetry of the obtained Hamiltonians is introduced and the mixed photon-magnetoexciton-acoustical plasmon states are discussed. The section 7 is devoted to the conclusions.

First of all we will describe the Landau quantization of the 2D heavy holes following the Ref.[19, 22].
\section{Landau quantization of the 2D heavy holes}
The full Landau-Rashba Hamiltonian for 2D heavy holes was discussed in Ref.[19] following the formulas (13)-(20). It can be expressed through the Bose-type creation and annihilation operators ${{a}^{\dagger }}$, $a$ acting on the Fock quantum states $\left| n \right\rangle =\frac{{{({{a}^{\dagger }})}^{n}}}{\sqrt{n!}}\left| 0 \right\rangle $, where $\left| 0 \right \rangle$ is the vacuum state of harmonic oscillator. The Hamiltonian has the form [22]
\begin{eqnarray}
\hat{H}_h &=& \hbar \omega_{ch} \left\lbrace \left[ \left( a^{\dag}a + \frac{1}{2}\right) + \delta \left( a^{\dag}a + \frac{1}{2}\right)^2 \right] \hat{I} \right. \nonumber \\
&& \left. + i \beta 2\sqrt{2} \left|\begin{array}{cc}
0 & (a^{\dag})^3\\
-a^3 & 0\end{array}\right| \right\rbrace; \hat{I} = \left|\begin{array}{cc}
1 & 0\\
0 & 1\end{array}\right| \label{eq:fullhameh}
\end{eqnarray}
with the denotations

\begin{eqnarray}
& \omega_{ch} = \frac{|e|H}{m_h c}; \: \delta = \frac{\left| \delta_h E_z\right| \hbar^4}{l^4 \hbar\omega_{ch}}; \: \beta = \frac{\beta_{h}E_{z}}{l^{3}\hbar\omega_{ch}}; \: l = \sqrt{\frac{\hbar c}{|e|H}}&. \label{eq:denotwdbl}
\end{eqnarray}
The parameter $\delta_h$ is not well known, what permits to consider different variants mentioned below.

The exact solutions of the Pauli-type Hamiltonian is described by the formulas (21)-(31) of the Ref.[19]. In more details they were described in Ref.[22], and have the spinor form
\begin{eqnarray}
& \hat{H}_h \left|\begin{array}{c}
f_{1}\\
f_{2}\end{array}\right| = E_{h} \left|\begin{array}{c}
f_{1}\\
f_{2}\end{array}\right|; \: f_1 = {\displaystyle \sum_{n=0}^{\infty} c_n \left| n \right \rangle}; \; f_2 = {\displaystyle \sum_{n=0}^{\infty} d_n \left| n \right \rangle}; & \nonumber \\
& {\displaystyle \sum_{n=0}^{\infty} |c_n|^2 + \sum_{n=0}^{\infty} |d_n|^2 = 1}. & \label{eq:solptham}
\end{eqnarray}
First three solutions depend only on one quantum number $m$ with the values $0, 1, 2$ as follows [6]

\begin{eqnarray}
& E_h(m=0) = \hbar \omega_{ch} \left( \frac12 + \delta \right); \: \Psi (m=0) = \left|\begin{array}{c}
\left| 0 \right \rangle \\
0 \end{array} \right|, & \nonumber \\
& E_h(m=1) = \hbar \omega_{ch} \left( \frac32 + 9 \delta \right); \: \Psi (m=1) = \left|\begin{array}{c}
\left| 1 \right \rangle \\
0 \end{array} \right|, & \nonumber \\
& E_h(m=2) = \hbar \omega_{ch} \left( \frac52 + 25 \delta \right); \: \Psi (m=2) = \left|\begin{array}{c}
\left| 2 \right \rangle \\
0 \end{array}\right|. & \label{eq:enhthreestates}
\end{eqnarray}
All another solutions with $m \ge 3$ depend on two quantum numbers $(m-5/2)$ and $(m+1/2)$ and have the general expression

\begin{eqnarray}
& \varepsilon_{h}^{\pm} (m - \frac52; m + \frac12) = \frac{E_{h}^{\pm} (m-5/2; m+1/2)}{\hbar \omega_{ch}} & \nonumber \\
& = (m-1) + \frac{\delta}{2} \left[ (2m+1)^2 + (2m-5)^2\right] & \nonumber \\
& \pm (\left( \frac32 + \frac{\delta}{2} \left[ (2m+1)^2 - (2m-5)^2\right] \right)^2 & \nonumber \\
& + 8 \beta^2 m(m-1)(m-2))^{1/2}, \: m \ge 3. &
\label{eq:exactsoldl}
\end{eqnarray}
The corresponding wave functions for $m=3$ and $m=4$ are

\begin{eqnarray}
& \Psi_{h}^{\pm} (m=3) = \left|\begin{array}{c}
c_{3} \left| 3 \right\rangle \\
d_{0} \left| 0 \right\rangle \end{array}\right| \: \textrm{and} \:
\Psi_{h}^{\pm} (m=4) = \left|\begin{array}{c}
c_{4} \left| 4 \right\rangle \\
d_{1} \left| 1 \right\rangle \end{array}\right|. & \label{eq:hwfspipar}
\end{eqnarray}
They depend on the coefficients $c_m$ and $d_{m-3}$, which obey to the equations

\begin{eqnarray}
& c_m \left( m + \frac12 + \delta (2m+1)^2 - \varepsilon_{h} \right) & \nonumber \\
& = -i \beta 2 \sqrt{2} \sqrt{m(m-1)(m-2)} d_{m-3}; & \nonumber \\
& d_{m-3} \left( m - \frac52 + \delta (2m-5)^2 - \varepsilon_{h} \right) & \nonumber \\
& = i \beta 2 \sqrt{2} \sqrt{m(m-1)(m-2)} c_{m}; & \nonumber \\
& |c_m|^2 + |d_{m-3}|^2 = 1.& \label{eq:displawhsyst}
\end{eqnarray}
There are two different solutions $\varepsilon_{h}^{\pm} (m)$ at a given value of $m \ge 3$ and two different pairs of the coefficients $(c_m^{\pm}, d_{m-3}^{\pm})$.
The dependences of the parameters $\omega_{ch}$, $\beta$ and $\delta$ on the electric and magnetic fields strengths may be represented for the GaAs-type quantum wells as follows $H = y \; \mathrm{T}$; $E_z = x \; \mathrm{kV/cm}$; $m_h = 0.25 m_0$; $\hbar \omega_{ch} = 0.4 y \; \mathrm{meV}$; $\beta = 1.062 \cdot 10^{-2} x \sqrt{y}$; $\delta = 10^{-4} C x y$ with unknown parameter $C$, which will be varied in more large interval of values. We cannot neglect the parameter $C$ putting it equal to zero, because in this case, as was argued in Ref.[19] formula (10), the lower spinor branch of the heavy hole dispersion law
\begin{eqnarray*}
& E_{h}^{-} (k_{||}) = \frac{\hbar^2 \vec{k}_{||}^2}{2m_h} - \left| \frac{\beta_h E_z}{2} \right| \left| \vec{k}_{||} \right|^3 & \label{eq:tspinbr}
\end{eqnarray*}
has an unlimited decreasing, deeply penetrating inside the semiconductor energy gap at great values of $\left| \vec{k}_{||} \right|$. To avoid this unphysical situation the positive quartic term $\left| \delta_hE_z \right| \vec{k}_{||}^4$ was added in the starting Hamiltonian. The new dependences were compared with the drawings calculated in the fig.2 of the Ref.[19] in the case $E_z = 10 \; \mathrm{kV/cm}$ and $C=10$. Four lowest Landau levels(LLLs) for heavy holes were selected in Ref.[19]. In addition to them in Ref.[22] were studied else three levels as follows
\begin{eqnarray}
& E_h(R_1) = E_h^{-} (\frac12, \frac72); \: E_h(R_2) = E_h(m=0); & \nonumber \\
& E_h(R_3) = E_h^{-} (\frac32, \frac92); \: E_h(R_4) = E_h(m=1); & \nonumber \\
& E_h(R_5) = E_h^{-} (\frac52, \frac{11}{2}); \: E_h(R_6) = E_h(m=2); & \nonumber \\
& E_h(R_7) = E_h^{-} (\frac72, \frac{13}{2}). &
\end{eqnarray}
Their dependencies on the magnetic field strength were represented in the figures 1 and 2 of the Ref.[22] at different parameters $x$ and $C$ and are reproduced below.
\begin{figure}[h]
\includegraphics[scale=0.55]{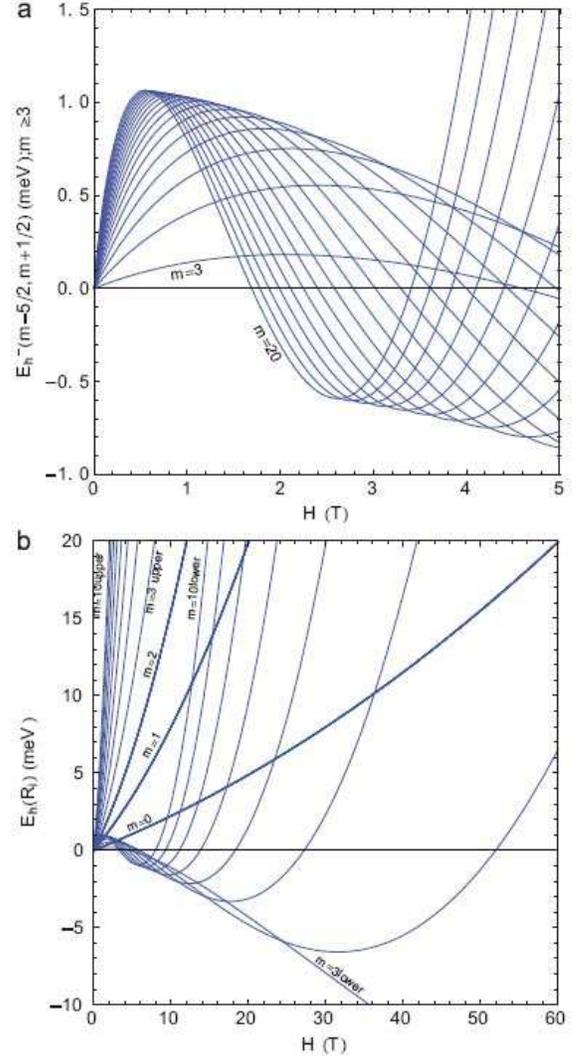}
\caption{a) The lower branches of the heavy hole Landau quantization levels $E_{h}^{-} (m-5/2; m+1/2)$ for $m \ge 3$ at the parameters $E_z = 10 \; \mathrm{kV/cm}$ and $C=5.5$; b) The general view of the all heavy hole Landau quantization levels with m=0,1,...,10 at the same parameters $E_z$ and $C$. They are reproduced from the fig.1 of the Ref.[22].}
\end{figure}

The general view of the lower branches $E_h^{-}(m-\frac52, m+\frac12)$ of the heavy hole Landau quantization levels with $m \ge 3$ as a functions of the magnetic field strength are represented in fig.1a following the formula (5). The upper branches have more simple monotonous behavior and are drawn in fig.1b together with some curves of the lower branches. All the lower branches in their initial parts have a linear increasing behavior up till they achieve the maximal values succeeded by the minimal values in the middle parts of their evolutions being transformed in the final quadratic increasing dependences. The values of the magnetic field strength corresponding to the minima and to the maxima decrease with the increasing of the number $m$. These peculiarities can be compared with the case of Landau quantization of the 2D electron in the biased bilayer graphene described in Ref.[18]. The last case is characterized by the initial dispersion law without parabolic part and by second order chirality terms. They both lead to dependences on the magnetic field strength for the lower dispersion branches with sharp initial decreasing parts and minimal values succeeded by the quadratic increasing behavior. The differences between the initial dispersion laws and chirality terms in two cases of bilayer graphene and of heavy holes lead to different intersections and degeneracies of the Landau levels.
	
The spinor-type envelope wave fynctions of the heavy holes in the coordinate representation look as [22]
\begin{eqnarray}
  |\psi_{h}(\varepsilon_{m},q;x,y)\rangle = \frac{e^{iqx}}{\sqrt{L_{x}}}\left\vert
\begin{array}{c}
\varphi_{m}(y+ql^{2}) \\
0%
\end{array}%
\right\vert ,\nonumber \\
  m=0,1,2, \\
|\psi_{h}(\varepsilon^{\pm}_{m},q;x,y)\rangle =  \frac{e^{iqx}}{\sqrt{L_{x}}}\left\vert
\begin{array}{c}
c^{\pm}_{m}\varphi_{m}(y+ql^{2}) \\
d^{\pm}_{m-3}\varphi_{m-3}(y+ql^{2})%
\end{array}%
\right\vert , \nonumber \\
  m \geq 3. \nonumber
\end{eqnarray}
The valence electrons in comparison with the holes are characterized by the opposite signs of he spin projections, wave vector and charge. The corresponding envelope wave functions can be obtained from the previous ones by the procedure
\begin{equation}
\left\vert \psi _{h}(\varepsilon _{m},q;x,y)\right\rangle =i\widehat{\sigma }_{y}\left\vert \psi _{h}(\varepsilon _{m},-q;x,y)\right\rangle ^{\ast }
\end{equation}
where ${{\hat{\sigma }}_{y}}$ is the Pauli matrix. In coordinate representation they are
\begin{eqnarray}
  |\psi_{v}(\varepsilon_{m},q;x,y)\rangle = \frac{e^{iqx}}{\sqrt{L_{x}}}\left\vert
\begin{array}{c}
0 \\
-\varphi^{*}_{m}(y-ql^{2})%
\end{array}%
\right\vert ,\nonumber \\
  m=0,1,2, \\
|\psi_{h}(\varepsilon^{\pm}_{m},q;x,y)\rangle =  \frac{e^{iqx}}{\sqrt{L_{x}}}\left\vert
\begin{array}{c}
d^{\pm *}_{m-3}\varphi^{*}_{m-3}(y-ql^{2}) \\
-c^{\pm *}_{m}\varphi^{*}_{m}(y-ql^{2})%
\end{array}%
\right\vert , \nonumber \\
  m \geq 3. \nonumber
\end{eqnarray}
To obtain the full valence electron Bloch wave functions the expressions (11) must be multiplied by the periodic parts. In the p-type valence band they have the form $\frac{1}{\sqrt{2}}({{U}_{v,p,x,q}}(x,y)\pm i{{U}_{v,p,y,q}}(x,y))$ and are characterized by the orbital momentum projections ${{M}_{v}}=\pm 1$ correspondingly. The hole orbital projections ${{M}_{h}}=-{{M}_{v}}$ have opposite signs in comparison with the valence electron. The full Bloch wave functions of the valence electrons now are characterized by a supplementary quantum number ${{M}_{v}}$ side by side with the previous ones ${{\varepsilon }_{m}}$, ${{\varepsilon }^{\pm }}_{m}$ and q as follows
\begin{eqnarray}
\left\vert \psi _{v}\left( M_{v},\varepsilon _{m},q;x,y\right) \right\rangle
&=&\frac{e^{iqx}}{\sqrt{L_{x}}}\frac{1}{\sqrt{2}}(U_{v,p,x,q}(\vec{r})
\nonumber \\
&&\pm iU_{v,p,y,q}(\vec{r}))\left\vert
\begin{array}{c}
0 \\
-\varphi _{m}^{\ast }(y-ql^{2})%
\end{array}%
\right\vert ;  \nonumber \\
m &=&0,1,2;M_{v}=\pm 1, \\
\left\vert \psi _{v}\left( M_{v},\varepsilon _{m}^{\pm },q;x,y\right)
\right\rangle  &=&\frac{e^{iqx}}{\sqrt{L_{x}}}\frac{1}{\sqrt{2}}(U_{v,p,x,q}(%
\vec{r})  \nonumber \\
&&\pm iU_{v,p,y,q}(\vec{r}))\left\vert
\begin{array}{c}
d_{m-3}^{\pm \ast }\varphi _{m-3}^{\ast }(y-ql^{2}) \\
-c_{m}^{\pm \ast }\varphi _{m}^{\ast }(y-ql^{2})%
\end{array}%
\right\vert ,  \nonumber \\
m &\geq &3,M_{v}=\pm 1  \nonumber
\end{eqnarray}
From this multitude of valence electron wave functions the more important of them are characterized by the values $\varepsilon _{m}^{-}$ with $m=3$ and 4, aw well as by $\varepsilon _{m}^{{}}$ with $m=0,1$. These four lowest hole energy levels being combined with two projections ${{M}_{h}}\pm 1$ form a set of 8 lowest hole states, which will be taken into account below.

Now for the completeness we will remember the main results obtained by Rashba [1] in the case of the electron conduction band. They are needed to obtain a full description of the 2D electron-hole pair and of a 2D magnetoexciton in the condition of the Landau quantization under the influence of the RSOC.

The lowest Landau level of the conduction electron in the presence of the RSOC was obtained in Ref.[1]:
\begin{eqnarray}
& \left| \psi_{\mathrm{e}}\left( R_{1},p;x_e,y_e \right)\right\rangle = \frac{e^{ipx_{e}}}{\sqrt{L_{x}}} \left|\begin{array}{c}
a_{0}\varphi_{0}(y_{\mathrm{e}})\\
b_{1}\varphi_{1}(y_{\mathrm{e}})\end{array}\right|; & \nonumber \\
& \varepsilon_{{\mathrm{e}}R_{1}} = 1 - \sqrt{\frac{1}{4} + 2\alpha^{2}}; \: |a_{0}|^{2} + |b_{1}|^{2} = 1 & \nonumber \\
& |a_{0}|^{2} = \frac{1}{1 + \frac{2\alpha^{2}}{\left[\frac{1}{2} + \sqrt{\frac{1}{4} + 2\alpha^{2}}\right]^{2}}}; \: |b_{1}|^{2} = \frac{2\alpha^{2}|a_{0}|^{2}}{\left[\frac{1}{2} + \sqrt{\frac{1}{4} + 2\alpha^{2}}\right]^{2}}.
\end{eqnarray}
The next electron level higher situated on the energy scale is characterized by the pure spin oriented state
\begin{eqnarray}
& \left| \psi_{\mathrm{e}}\left( R_{2},p;x_{\mathrm{e}},y_{\mathrm{e}} \right)\right\rangle = \frac{e^{ipx_{e}}}{\sqrt{L_{x}}} \left|\begin{array}{c}
0\\
\varphi_{0}(y_{\mathrm{e}})\end{array}\right|; \: \varepsilon_{{\mathrm{e}}R_{2}} = \frac12. &
\end{eqnarray}
Two lowest Landau levels~(LLLs) for conduction electron are characterized by the values $m_{\mathrm{e}} = 0.067 m_0$, $\hbar \omega_{\mathrm{ce}} = 1.49 \; \mathrm{meV}  \cdot y$ and parameter $\alpha = 8 \cdot 10^{-3} x/ \sqrt{y}$. They are denoted as

\begin{eqnarray}
& E_{\mathrm{e}}(R_{1}) = \hbar \omega_{\mathrm{ce}} \left( 1 - \sqrt{\frac{1}{4} + 2\alpha^{2}} \right); \nonumber \\
& E_{\mathrm{e}}(R_{2}) = \hbar \omega_{\mathrm{ce}} \frac12. &
\end{eqnarray}

The lowest Landau energy level for electron $E_{\mathrm{e}}(R_1)$ has a nonmonotonous anomalous dependence on the magnetic field strength near the point $H=0 \: \mathrm{T}$. It is due to the singular dependence of the RSOC parameter $\alpha^2 = 6.4 \cdot 10^{-5} x^2/y$, which is compensated in the total energy level expression by the factor $\hbar \omega_{\mathrm{ce}}$ of the cyclotron energy, where $\hbar \omega_{\mathrm{ce}} = 1.49 \: y$~meV. The second electron Landau energy level has a simple linear dependence on $H$.

The full Bloch wave functions for conduction electrons including their s-type periodic parts look as
\begin{eqnarray}
\left\vert \psi _{c}\left( s,R_{1},p;x,y\right) \right\rangle  &=&\frac{%
e^{ipx}}{\sqrt{L_{x}}}U_{c,s,p}(\vec{r})\left\vert
\begin{array}{c}
a_{0}\varphi _{0}(y-pl^{2}) \\
b_{1}\varphi _{1}(y-pl^{2})%
\end{array}%
\right\vert ,  \nonumber \\
\left\vert \psi _{c}\left( s,R_{2},p;x,y\right) \right\rangle  &=&\frac{%
e^{ipx}}{\sqrt{L_{x}}}U_{c,s,p}(\vec{r})\left\vert
\begin{array}{c}
0 \\
\varphi _{m}(y-pl^{2})%
\end{array}%
\right\vert .
\end{eqnarray}
Two lowest Rashba-type states for conduction electron will be combined with eight LLLs for heavy holes and with the corresponding states of the valence electrons. The e-h pair will be characterized by 16 states. Heaving the full set of the electron Bloch wave functions in conduction and in the valence bands one can construct the Hamiltonian describing in second quantization representation the Coulomb electron-electron interaction as well as the electron-radiation interaction. These tow tasks will be described in the next sections of our review paper. The results obtained earlier in the Ref.[19, 22] taking into account only 8 e-h states will be supplemented below.

\section{Electron-radiation interaction in the presence of the Rashba spin-orbit coupling}
In the Ref.[17, 21] the Hamiltonian of the electron-radiation interaction in the second quantization representation for the case of two-dimensional(2D) coplanar electron-hole(e-h) system in a strong perpendicular magnetic field was discussed. The $s$-type conduction-band electrons with spin projections $s_z = \pm 1/2$ along the magnetic field direction and the heavy holes with the total momentum projections $j_z = \pm 3/2$ in the $p$-type valence band were taken into account. Their orbital Bloch wave functions are similar to $(x \pm iy)$ expressions with the orbital momentum projections $M= \pm 1$ on the same selected direction. The Landau quantization of the 2D electrons and holes was described in the Landau gauge with oscillator type motion in one in-plane direction characterized by the quantum numbers $n_e$ and $n_h$ and with the free translational motion described by the uni-dimensional(1D) wave numbers $p$ and $q$ in another in-plane direction perpendicular to the previous one. The electron and hole creation and annihilation operators $a^{+}_{s_z, n_e, p}$, $a_{s_z, n_e, p}$, and $b^{+}_{j_z, n_h, q}$, $b_{j_z, n_h, q}$ were introduced correspondingly. The Zeeman effect and the Rashba spin-orbit coupling in Refs [17, 21] were not taken into account.
	
The electrons and holes have a free orbital motion on the surface of the layer with the area $S$ and are completely confined in $\vec{a}_3$ direction. The degeneracy of their Landau levels equals to $N = S/(2 \pi l_0^2)$, where $l_0$ is the magnetic length. In contrast, the photons were supposed to move in any direction in the three-dimensional(3D) space with the wave vector $\vec{k}$ arbitrary oriented as regards the 2D layer as it is represented in the Fig.2 reproduced from the Ref.[17]. There are three unit vectors $\vec{a}_1$, $\vec{a}_2$, $\vec{a}_3$, the first two being in-plane oriented whereas the third $\vec{a}_3$ is perpendicular to the layer. We will use the 3D and 2D wave vectors $\vec{k}$ and $\vec{k}_{||}$ and will introduce the circular polarization vectors $\vec{\sigma}_{M}$ for the valence electrons, heavy holes and magnetoexcitons as follows
\begin{figure}[h]
\includegraphics[scale=0.35]{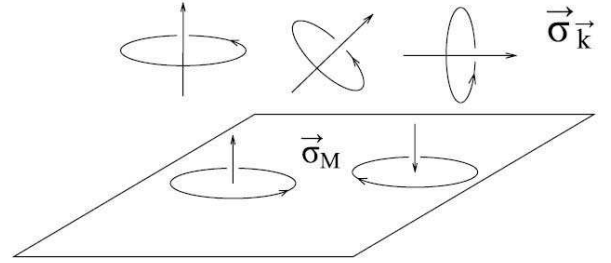}
\caption{The reciprocal orientations of the circularly polarized vectors $\vec{\sigma}_{\vec{k}}$ and $\vec{\sigma}_{M}$, reproduced from the Ref.[17].}
\end{figure}
\begin{eqnarray}
  \vec{k}=\vec{k}_{||}+\vec{a}_{3}k_{z}, \nonumber \\
  \vec{k}_{||}=\vec{a}_{1}k_{x}+\vec{a}_{2}k_{y}, \nonumber \\
  \vec{\sigma}_{M}=\frac{1}{\sqrt{2}}(\vec{a}_{1}\pm i\vec{a}_{2}), M=\pm 1.
\end{eqnarray}
The photons are characterized by two linear vectors $\vec{e}_{k,j}$ or by two circular polarization vectors $\vec{\sigma}_{\vec{k}}^{\pm}$ obeying the transversality conditions:
\begin{eqnarray}
\vec{\sigma }_{{\vec{k}}}^{\pm }=\frac{1}{\sqrt{2}}({{{\vec{e}}}_{\vec{k},1}}\pm i{{{\vec{e}}}_{\vec{k},2}}), \nonumber \\
 ({{{\vec{e}}}_{k,j}}\cdot \vec{k})=0; j=1,2.
\end{eqnarray}
The photon creation and annihilation operators can be introduced in two different polarizations as follows
\begin{eqnarray}
{{C}_{\vec{k},\pm }}=\frac{1}{\sqrt{2}}\left( {{C}_{\vec{k},1}}\pm i{{C}_{\vec{k},2}} \right), \nonumber \\
 {{\left( {{C}_{\vec{k},\pm }} \right)}^{\dagger }}=\frac{1}{\sqrt{2}}\left( C_{\vec{k},1}^{\dagger }\mp i C_{\vec{k},2}^{\dagger } \right), \nonumber \\
\sum^{2}_{j=1}{{\vec{e}}_{\vec{k},j}}{{C}_{\vec{k},j}}={{C}_{\vec{k},-}}\vec{\sigma }_{{\vec{k}}}^{+}+{{C}_{\vec{k},+}}\vec{\sigma }_{{\vec{k}}}^{-}, \nonumber \\
\sum^{2}_{j=1}{{\vec{e}}_{\vec{k},j}}C_{\vec{k},j}^{\dagger }={{\left( {{C}_{\vec{k},-}} \right)}^{\dagger }}\vec{\sigma }_{{\vec{k}}}^{-}+{{\left( {{C}_{\vec{k},+}} \right)}^{\dagger }}\vec{\sigma }_{{\vec{k}}}^{+}.
\end{eqnarray}
The reciprocal orientations of the circular polarizations $\vec{\sigma}_{\vec{k}}^{\pm}$ and $\vec{\sigma}_{M}$ will determine the values of the scalar products $(\vec{\sigma}_{\vec{k}}^{\pm} \cdot \vec{\sigma}_{M}^{\ast})$. The electron-radiation interaction describing only the band-to-band quantum transitions with the participation of the e-h pairs in the presence of a strong perpendicular magnetic field was obtained in Ref.[17] and can be used as initial expression for obtaining the interaction of 2D magnetoexcitons with the electromagnetic field.

These results will be generalized below taking into account the Rashba spin-orbit coupling, what means the use of the spinor-type wave functions (12) and (16) instead of the scalar ones [17, 21]. The Hamiltonian looks as
\begin{widetext}
\begin{eqnarray}
H_{e-rad}=(-\frac{e}{m_{0}})\sum_{\vec{k}(\vec{k}_{||}, k_{z})} \sqrt{\frac{2\pi \hbar}{V\omega_{\vec{k}}}} \sum_{i=1,2} \sum_{M_{v}=\pm 1} \sum_{\varepsilon=\varepsilon_{m},\varepsilon^{-}_{m}} \sum_{g,q}([C_{\vec{k},-}\vec{\sigma}^{+}_{\vec{k}}+C_{\vec{k},+}\vec{\sigma}^{-}_{\vec{k}}]\cdot  \nonumber \\
\cdot[\vec{P}(c,R_{i},g;v,M_{v},\varepsilon,q;\vec{k})a^{\dag}_{c,R_{i},g}a_{v,M_{v},\varepsilon,q}+\vec{P}(v,M_{v},\varepsilon,q;c,R_{i},g;\vec{k})a^{\dag}_{v,M_{v},\varepsilon,q}a_{c,R_{i},g}]+ \nonumber \\
+[(C_{\vec{k},-})^{\dag}\vec{\sigma}^{-}_{\vec{k}}+(C_{\vec{k},+})^{\dag}\vec{\sigma}^{+}_{\vec{k}}]\cdot[\vec{P}(c,R_{i},g;v,M_{v},\varepsilon,q;-\vec{k})\times \nonumber \\
\times a^{\dag}_{c,R_{i},g}a_{v,M_{v},\varepsilon,q}+\vec{P}(v,M_{v},\varepsilon,q;c,R_{i},g;-\vec{k})a^{\dag}_{v,M_{v},\varepsilon,q}a_{c,R_{i},g}])
\end{eqnarray}
\end{widetext}
The matrix elements will be discussed below. One of them has the form
\begin{eqnarray}
&\vec{P}(c,{{R}_{1}},g;v,{{M}_{v}},{{\varepsilon }^{-}}_{m},q;\vec{k})= &\nonumber \\
&=\int{{{d}^{2}}\vec{r}}\left\langle  {{{\hat{\psi }}}_{c,{{R}_{i}},g}}(\vec{r}) \right|{{e}^{i\vec{k}\vec{r}}}\hat{\vec{p}}\left| {{{\hat{\psi }}}_{v,{{M}_{v}},{{\varepsilon }^{-}}_{m},q}}(\vec{r}) \right\rangle = &\nonumber \\
&=\frac{a_{0}^{*}d_{m-3}^{-*}}{\sqrt{2}{{L}_{x}}}\int{{{d}^{2}}\vec{r}}U_{c,s,g}^{*}(\vec{r}){{e}^{-igx}}\varphi _{0}^{*}(y-g{{l}^{2}})\times & \nonumber \\
&\times {{e}^{i\vec{k}\vec{r}}}({{{\vec{a}}}_{1}}{{{\hat{p}}}_{x}}+{{{\vec{a}}}_{2}}{{{\hat{p}}}_{y}})({{U}_{v,p,x,q}}(\vec{r})\pm i{{U}_{v,p,y,q}}(\vec{r}))\times  &\nonumber \\
&\times {{e}^{iqx}}\varphi _{m-3}^{*}(y-p{{l}^{2}})-& \nonumber \\
&-\frac{b_{1}^{*}c_{m}^{-*}}{\sqrt{2}{{L}_{x}}}\int{{{d}^{2}}\vec{r}}U_{c,s,g}^{*}(\vec{r}){{e}^{-igx}}\times & \nonumber \\
&\times \varphi _{1}^{*}(y-g{{l}^{2}}){{e}^{i\vec{k}\vec{r}}}({{{\vec{a}}}_{1}}{{{\hat{p}}}_{x}}+{{{\vec{a}}}_{2}}{{{\hat{p}}}_{y}})\times  &\nonumber \\
&\times ({{U}_{v,p,x,q}}(\vec{r})\pm i{{U}_{v,p,y,q}}(\vec{r})){{e}^{iqx}}\varphi _{m}^{*}(y-p{{l}^{2}})&
\end{eqnarray}
One can represent the 2D coordinate vector $\vec{r}$ as a sum $\vec{r}=\vec{R}+\vec{\rho }$ of a lattice point vector $\vec{R}$ and of a small vector $\vec{\rho }$ changing inside the unit lattice cell with lattice constant ${{a}_{0}}$ and volume ${{v}_{0}}=a_{0}^{3}$. Any 2D semiconductor layer has at least the minimal width ${{a}_{0}}$ and the periodic parts ${{U}_{nk}}(\vec{r})$ are determined inside the elementary lattice cell. The periodic parts ${{U}_{nk}}(\vec{r})$ do not depend on $\vec{R}$ because ${{U}_{nk}}(\vec{R}+\vec{\rho })={{U}_{nk}}(\vec{\rho })$. From another side the envelope functions ${{\varphi }_{n}}(\vec{r})$ describing the Landau quantization have a spread of the order of magnetic length ${{l}_{0}}$ which is much greater than ${{a}_{0}}$ (${{l}_{0}}>>{{a}_{0}}$). It means that they practically do not depend on $\vec{\rho }$, i.e. ${{\varphi }_{n}}(\vec{R}+\vec{\rho })\cong {{\varphi }_{n}}(\vec{R})$. The matrix elements (21) contains some functions which do not depend on $\vec{R}$ and another ones, which do not depend on $\vec{\rho }$. Only the plane wave ${{e}^{i\vec{k}\vec{r}}}={{e}^{i\vec{k}\vec{R}+i\vec{k}\vec{\rho }}}$ contains both of them. The derivative $\frac{\partial }{\partial \vec{r}}$ acts in the same manner on the functions ${{U}_{nk}}(\vec{\rho })$ and ${{\varphi }_{n}}(\vec{R})$ because $\vec{R}$ and $\vec{\rho }$ are the components of $\vec{r}$. These properties suggested to transform the 2D integral on the variable $\vec{r}$ in two separate integrals on the variables $\vec{R}$ and $\vec{\rho }$ as follows
\begin{eqnarray}
\int{{{d}^{2}}\vec{r}}A(\vec{R})B(\vec{\rho })=\sum\limits_{{\vec{R}}}{A(\vec{R})}\int{{{d}^{2}}\vec{\rho }}B(\vec{\rho })= \nonumber \\
=\sum\limits_{{\vec{R}}}{A(\vec{R})}a_{0}^{2}\frac{1}{{{v}_{0}}}\int{{{d}^{3}}\vec{\rho }}B(\vec{\rho })= \nonumber \\
=\int{{{d}^{2}}\vec{R}}A(\vec{R})\frac{1}{{{v}_{0}}}\int\limits_{{{v}_{0}}}{d\vec{\rho }}B(\vec{\rho })
\end{eqnarray}
Here the small value $a_{0}^{2}$ is substituted by the infinitesimal differential value ${{d}^{2}}\vec{R}$ because $A(\vec{R})$ is a smooth function on $\vec{R}$. The integrals on the volume ${{v}_{0}}$ of the elementary lattice cell contain the quickly oscillating periodic parts $U_{c,s,g}^{{}}(\vec{\rho })$ and ${{U}_{v,p,i,q}}(\vec{\rho })$ belonging to s-type conduction band and to p-type valence band. They have different parities and obey to selection rules
\begin{eqnarray}
&\frac{1}{{{v}_{0}}}\int\limits_{{{v}_{0}}}{d\vec{\rho }}U_{c,s,g}^{*}(\vec{\rho }){{e}^{i{{k}_{y}}{{\rho }_{y}}}}{{U}_{v,p,i,q}}(\vec{\rho })=0, &\nonumber \\
&i,j=x,y,& \nonumber \\
&\frac{1}{{{v}_{0}}}\int\limits_{{{v}_{0}}}{d\vec{\rho }}U_{c,s,g}^{*}(\vec{\rho }){{e}^{i{{k}_{y}}{{\rho }_{y}}}}\frac{\partial }{\partial {{\rho }_{i}}}{{U}_{v,p,j,q}}(\vec{\rho })=0, &\nonumber \\
&\text{if }i\ne j.&
\end{eqnarray}
The case $i=j$ is different from zero and gives rise to the expression
\begin{equation}
\frac{1}{{{v}_{0}}}\int\limits_{{{v}_{0}}}{d\vec{\rho }}U_{c,s,g}^{*}(\vec{\rho }){{e}^{i{{k}_{y}}{{\rho }_{y}}}}\frac{\partial }{\partial {{\rho }_{i}}}{{U}_{v,pi,g-{{k}_{x}}}}(\vec{\rho })={{P}_{cv}}({{\vec{k}}_{||}},g)
\end{equation}
The last integral in the zeroth approximation is of the allowed type in the definition of Elliott [26, 27] and can be considered as a constant ${{P}_{cv}}({{\vec{k}}_{||}},g)\approx {{P}_{cv}}(0)$ which does not depend on the wave vectors ${{\vec{k}}_{||}}$ and $g$. Due to these selection rules the derivatives $\partial /\partial \vec{r}$ in the expression (21) must be taken only from the periodic parts ${{U}_{v,p,i,q}}(\vec{\rho })$ because all another integrals vanish.
The integration on the variable ${{R}_{x}}$ engages only the plane wave functions and gives rise to the selection rule for the 1D wave numbers $g,q,{{k}_{x}}$ as follows
\begin{equation}
\frac{1}{{{L}_{x}}}\int{{}}{{e}^{i{{R}_{x}}(q-g+{{k}_{x}})}}=\frac{2\pi }{{{L}_{x}}}\delta (q-g+{{k}_{x}})={{\delta }_{kr}}(q,g-{{k}_{x}})
\end{equation}
The integral on the variable ${{R}_{y}}$ engages only the Landau quantization functions ${{\varphi }_{n}}({{R}_{y}})$ and gives rise to the third selection rule concerning the numbers ${{n}_{e}}$ and ${{n}_{h}}$ of the Landau levels for electrons and holes. It looks as
\begin{eqnarray}
&\int\limits_{-\infty }^{\infty }{d{{R}_{y}}}\varphi _{{{n}_{e}}}^{*}({{R}_{y}}-g l_{0}^{2})\varphi _{{{n}_{h}}}^{*}({{R}_{y}}-(g-{{k}_{x}})l_{0}^{2}){{e}^{i{{k}_{y}}{{R}_{y}}}}= &\nonumber \\
&={{e}^{i{{k}_{y}}g l_{0}^{2}}}{{e}^{-i\frac{{{k}_{x}}{{k}_{y}}}{2}l_{0}^{2}}}\tilde{\phi }({{n}_{e}},{{n}_{h}};{{{\vec{k}}}_{||}})&
\end{eqnarray}
where
\begin{eqnarray}
\tilde{\phi }({{n}_{e}},{{n}_{h}};{{{\vec{k}}}_{||}})=\int\limits_{-\infty }^{\infty }{dy}\varphi _{{{n}_{e}}}^{{}}(y-\frac{{{k}_{x}}l_{0}^{2}}{2})\varphi _{{{n}_{h}}}^{{}}(y+\frac{{{k}_{x}}l_{0}^{2}}{2}){{e}^{i{{k}_{y}}y}}, \nonumber \\
{{e}^{-i\frac{{{k}_{x}}{{k}_{y}}}{2}l_{0}^{2}}}\tilde{\phi }({{n}_{e}},{{n}_{h}};{{{\vec{k}}}_{||}})=\phi ({{n}_{e}},{{n}_{h}};{{{\vec{k}}}_{||}}). \nonumber
\end{eqnarray}
Here we took into account that the Landau quantization functions ${{\varphi }_{n}}(y)$ are real.
This selection rule coincides with the formula (24) in the absence of the RSOC and its interpretation remains the same. Once again one can underlain that during the dipole-active band-to-band quantum transition the numbers of the Landau levels in the initial starting band as well as in the final arriving band coincide i.e. ${{n}_{e}}={{n}_{h}}$. It is true in the equal manner in the absence as well as in the presence of the RSOC.
Three separate integrations on $\vec{\rho },{{R}_{x}}\text{ and }{{R}_{y}}$ taking into account the selection rules (23), (25), (26) lead to the expression
\begin{eqnarray}
&\vec{P}(c,{{R}_{1}},g;v,{{M}_{v}},\varepsilon _{m}^{-},q;{{{\vec{k}}}_{||}})=& \nonumber \\
&={{\delta }_{kr}}(q,g-{{k}_{x}}){{{\vec{\sigma }}}_{{{M}_{v}}}}{{P}_{cv}}(0){{e}^{i{{k}_{y}}g l_{0}^{2}}}\times & \nonumber \\
&\times {{e}^{-i\frac{{{k}_{x}}{{k}_{y}}}{2}l_{0}^{2}}}[a_{0}^{*}d_{m-3}^{-*}\tilde{\phi }(0,m-3;{{{\vec{k}}}_{||}})-b_{1}^{*}c_{m}^{-*}\tilde{\phi }(1,m;{{{\vec{k}}}_{||}})], &\nonumber\\
& m\geq 3&
\end{eqnarray}
Here the vectors of the circular polarizations ${{\vec{\sigma }}_{{{M}_{v}}}}$ describing the valence electron states were introduced following the formula (17). One can introduce also the vectors of the heavy hole circular polarizations ${{\vec{\sigma }}_{{{M}_{h}}}}$ in the form
\begin{eqnarray}
{{{\vec{\sigma }}}_{{{M}_{v}}}}=\frac{1}{\sqrt{2}}({{{\vec{a}}}_{1}}\pm i{{{\vec{a}}}_{2}}),{{M}_{v}}=\pm 1, \nonumber \\
{{{\vec{\sigma }}}_{{{M}_{h}}}}={{{\vec{\sigma }}}^{*}}_{{{M}_{v}}}=\frac{1}{\sqrt{2}}({{{\vec{a}}}_{1}}\mp i{{{\vec{a}}}_{2}}),{{M}_{h}}=\mp 1
\end{eqnarray}
The magnetoexciton states are characterized by the quantum numbers ${{M}_{h}}$, ${{R}_{i}}$, $\varepsilon $ and by the wave vectors ${{\vec{k}}_{||}}$.
The general expressions for the matrix elements are
\begin{eqnarray}
&\vec{P}(c,{{R}_{i}},g;v,{{M}_{v}},\varepsilon ,q;{{{\vec{k}}}_{||}})=& \nonumber \\
&={{\delta }_{kr}}(q,g-{{k}_{x}}){{{\vec{\sigma }}}_{{{M}_{v}}}}{{P}_{cv}}(0){{e}^{i{{k}_{y}}g l_{0}^{2}}}T(c{{R}_{i}},\varepsilon ;{{{\vec{k}}}_{||}}){{e}^{-i\frac{{{k}_{x}}{{k}_{y}}}{2}l_{0}^{2}}},& \nonumber \\
&i=1,2,\text{ }{{\text{M}}_{v}}=\pm 1,\text{ }\varepsilon ={{\varepsilon }_{m}},& \nonumber \\
&\text{with }m=0,1,2,\text{and }\varepsilon =\varepsilon _{m}^{-},\text{with }m\ge 3&
\end{eqnarray}
The coefficients $T(c{{R}_{i}},\varepsilon ;{{\vec{k}}_{||}})$ have the forms
\begin{eqnarray}
&T({{R}_{1}},\varepsilon _{m}^{-};{{{\vec{k}}}_{||}})=[a_{0}^{*}d_{m-3}^{-*}\tilde{\phi }(0,m-3;{{{\vec{k}}}_{||}})-b_{1}^{*}c_{m}^{-*}\tilde{\phi }(1,m;{{{\vec{k}}}_{||}})],& \nonumber\\
&m\ge 3,&\nonumber \\
&T({{R}_{1}},{{\varepsilon }_{m}};{{{\vec{k}}}_{||}})=[-b_{1}^{*}\tilde{\phi }(1,m;{{{\vec{k}}}_{||}})], &\nonumber\\
&m=0,1,2, &\nonumber\\
&T({{R}_{2}},\varepsilon _{m}^{-};{{{\vec{k}}}_{||}})=[-c_{m}^{-*}\tilde{\phi }(0,m;{{{\vec{k}}}_{||}})],& \nonumber\\
&m\ge 3,&\nonumber \\
&T({{R}_{2}},\varepsilon _{m}^{{}};\vec{k})=[-\tilde{\phi }(0,m;{{{\vec{k}}}_{||}})], &\nonumber\\
&m=0,1,2&
\end{eqnarray}
Another matrix elements can be calculated in the similar way. They are
\begin{eqnarray}
&\vec{P}(v,{{M}_{v}},\varepsilon ,q;c,{{R}_{i}},g;-{{{\vec{k}}}_{||}})= &\nonumber \\
& ={{{\vec{P}}}^{*}}(c,{{R}_{i}},g;v,{{M}_{v}},\varepsilon ,q;{{{\vec{k}}}_{||}})= & \nonumber\\
& ={{\delta }_{kr}}(q,g-{{k}_{x}})\vec{\sigma }_{{{M}_{v}}}^{*}P_{cv}^{*}(0){{e}^{-i{{k}_{y}}g l_{0}^{2}}}{{e}^{i\frac{{{k}_{x}}{{k}_{y}}}{2}l_{0}^{2}}}{{T}^{*}}({{R}_{i}},\varepsilon ;{{{\vec{k}}}_{||}}), & \nonumber \\
& \vec{P}(c,{{R}_{i}},g;v,{{M}_{v}},\varepsilon ,q;-{{{\vec{k}}}_{||}})=& \nonumber \\
& -{{\delta }_{kr}}(q,g+{{k}_{x}})\vec{\sigma }_{{{M}_{v}}}^{{}}P_{cv}^{{}}(0){{e}^{-i{{k}_{y}}g l_{0}^{2}}}{{e}^{-i\frac{{{k}_{x}}{{k}_{y}}}{2}l_{0}^{2}}}T({{R}_{i}},\varepsilon ;-{{{\vec{k}}}_{||}}),& \nonumber  \\
& \vec{P}(v,{{M}_{v}},\varepsilon ,q;c,{{R}_{i}},g;{{{\vec{k}}}_{||}})= &  \\
& ={{\delta }_{kr}}(q,g+{{k}_{x}})\vec{\sigma }_{{{M}_{v}}}^{*}P_{cv}^{*}(0){{e}^{i{{k}_{y}}g l_{0}^{2}}}{{e}^{i\frac{{{k}_{x}}{{k}_{y}}}{2}l_{0}^{2}}}{{T}^{*}}({{R}_{i}},\varepsilon ;-{{{\vec{k}}}_{||}})& \nonumber
\end{eqnarray}
They permit to calculate the electron operator parts in the Hamiltonian (20) and to express them through the magnetoexciton creation and annihilation operators determined as follows
\begin{equation}
\hat{\psi }_{ex}^{\dagger }({{\vec{k}}_{||}},{{M}_{h}},{{R}_{i}},\varepsilon )=\frac{1}{\sqrt{N}}\sum\limits_{t}{{}}{{e}^{i{{k}_{y}}tl_{0}^{2}}}a_{{{R}_{i}},\frac{{{k}_{x}}}{2}+t}^{\dagger }b_{{{M}_{h}},\varepsilon ,\frac{{{k}_{x}}}{2}-t}^{\dagger }
\end{equation}
Here the electron and hole creation and annihilation operator where introduced
\begin{eqnarray}
& {{a}_{{{R}_{i}},g}}={{a}_{c,{{R}_{i}},g}}, & \nonumber \\
& {{a}_{v,{{M}_{v}},\varepsilon ,q}}=b_{-{{M}_{h}},\varepsilon ,-q}^{\dagger } &
\end{eqnarray}
Here we have supposed that the Coulomb electron-hole interaction leading to the formation of the magnetoexciton is greater than the magnetoexciton-photon interaction leading to the formation of the magnetopolariton. It means that the ionization potential of the magnetoexciton ${{I}_{l}}$ is greater than the Rabi energy $\hbar \left| {{\omega }_{R}} \right|\approx \frac{\left| e \right|}{{{m}_{0}}{{l}_{0}}}\left| {{P}_{cv}}(0) \right|\sqrt{\frac{\hbar }{{{L}_{z}}{{\omega }_{{\vec{k}}}}}}$. It was determined in Ref.[21].

The existence of the phase factors of the type ${{e}^{\pm i{{k}_{y}}gl_{0}^{2}}}$ in the expressions (29) and (31) similar with that entering in the definitions of the magnetoexciton creation operators permits to obtain the expressions
\begin{eqnarray}
&\sum\limits_{q,g}{{}}\vec{P}(c,{{R}_{i}},g;v,{{M}_{v}},\varepsilon ,q;{{{\vec{k}}}_{||}})a_{c,{{R}_{i}},g}^{\dagger }{{a}_{v,{{M}_{v}},\varepsilon ,q}}=& \nonumber \\
& =\vec{\sigma }_{{{M}_{h}}}^{*}{{P}_{cv}}(0)T({{R}_{i}},\varepsilon ;{{{\vec{k}}}_{||}})\sqrt{N}\hat{\psi }_{ex}^{\dagger }({{{\vec{k}}}_{||}},{{M}_{h}},{{R}_{i}},\varepsilon ),& \nonumber \\
& \sum\limits_{q,g}{{}}\vec{P}(v,{{M}_{v}},\varepsilon ,q;c,{{R}_{i}},g;-{{{\vec{k}}}_{||}})a_{v,{{M}_{v}},\varepsilon ,q}^{\dagger }{{a}_{c,{{R}_{i}},g}}=& \nonumber \\
& =\vec{\sigma }_{{{M}_{h}}}^{{}}{{P}^{*}}_{cv}(0){{T}^{*}}({{R}_{i}},\varepsilon ;{{{\vec{k}}}_{||}})\sqrt{N}\hat{\psi }_{ex}^{{}}({{{\vec{k}}}_{||}},{{M}_{h}},{{R}_{i}},\varepsilon ),& \nonumber \\
& \sum\limits_{q,g}{{}}\vec{P}(c,{{R}_{i}},g;v,{{M}_{v}},\varepsilon ,q;-{{{\vec{k}}}_{||}})a_{c,{{R}_{i}},g}^{\dagger }{{a}_{v,{{M}_{v}},\varepsilon ,q}}=& \nonumber \\
& =\vec{\sigma }_{{{M}_{h}}}^{*}{{P}_{cv}}(0)T({{R}_{i}},\varepsilon ;-{{{\vec{k}}}_{||}})\sqrt{N}\hat{\psi }_{ex}^{\dagger }(-{{{\vec{k}}}_{||}},{{M}_{h}},{{R}_{i}},\varepsilon ),& \nonumber \\
& \sum\limits_{q,g}{{}}\vec{P}(v,{{M}_{v}},\varepsilon ,q;c,{{R}_{i}},g;{{{\vec{k}}}_{||}})a_{v,{{M}_{v}},\varepsilon ,q}^{\dagger }{{a}_{c,{{R}_{i}},g}}=& \\
& =\vec{\sigma }_{{{M}_{h}}}^{{}}{{P}^{*}}_{cv}(0){{T}^{*}}({{R}_{i}},\varepsilon ;-{{{\vec{k}}}_{||}})\sqrt{N}\hat{\psi }_{ex}^{{}}(-{{{\vec{k}}}_{||}},{{M}_{h}},{{R}_{i}},\varepsilon ),& \nonumber
\end{eqnarray}
In the Ref.[21] the Hamiltonian of the electron-radiation interaction was deduced in the absence of the RSOC. In its presence the mentioned Hamiltonian also can be expressed in compact form through the photon and magnetoexciton creation and annihilation operators. As earlier we introduced the values $N=S/2\pi l_{0}^{2}$, $V=S{{L}_{z}}$, where ${{L}_{z}}$ is the size of the 3D space in direction perpendicular to the layer. In the case of microcavity ${{L}_{z}}$ equals to the cavity length ${{L}_{c}}$. The electron-radiation interaction has the form
\begin{widetext}
\begin{eqnarray}
{{{\hat{H}}}_{e-rad}}=\left( -\frac{e}{{{m}_{0}}{{l}_{0}}} \right)\sum\limits_{\vec{k}({{{\vec{k}}}_{||}},{{k}_{z}})}{\sum\limits_{{{M}_{h}}=\pm 1}{\sum\limits_{i=1,2}^{{}}{\sum\limits_{\varepsilon ={{\varepsilon }_{m}},\varepsilon _{m}^{-}}{\sqrt{\frac{\hbar }{{{L}_{z}}{{\omega }_{{\vec{k}}}}}}}}}}\times  \nonumber\\
\times \{{{P}_{cv}}(0)T\left( {{R}_{i}},\varepsilon ,{{{\vec{k}}}_{||}} \right)[{{C}_{\vec{k},-}}(\vec{\sigma }_{{\vec{k}}}^{+}\cdot \vec{\sigma }_{{{M}_{h}}}^{*})+{{C}_{\vec{k},+}}(\vec{\sigma }_{{\vec{k}}}^{-}\cdot \vec{\sigma }_{{{M}_{h}}}^{*})]\hat{\psi }_{ex}^{\dagger }({{{\vec{k}}}_{||}},{{M}_{h}},{{R}_{i}},\varepsilon )+ \nonumber\\
+{{P}^{*}}_{cv}(0){{T}^{*}}\left( {{R}_{i}},\varepsilon ,{{{\vec{k}}}_{||}} \right)[{{\left( {{C}_{\vec{k},-}} \right)}^{\dagger }}(\vec{\sigma }_{{\vec{k}}}^{-}\cdot \vec{\sigma }_{{{M}_{h}}}^{{}})+{{\left( {{C}_{\vec{k},+}} \right)}^{\dagger }}(\vec{\sigma }_{{\vec{k}}}^{+}\cdot \vec{\sigma }_{{{M}_{h}}}^{{}})]\hat{\psi }_{ex}^{{}}({{{\vec{k}}}_{||}},{{M}_{h}},{{R}_{i}},\varepsilon )+ \nonumber\\
+{{P}_{cv}}(0)T\left( {{R}_{i}},\varepsilon ,-{{{\vec{k}}}_{||}} \right)[{{\left( {{C}_{\vec{k},-}} \right)}^{\dagger }}(\vec{\sigma }_{{\vec{k}}}^{-}\cdot \vec{\sigma }_{{{M}_{h}}}^{*})+{{\left( {{C}_{\vec{k},+}} \right)}^{\dagger }}(\vec{\sigma }_{{\vec{k}}}^{+}\cdot \vec{\sigma }_{{{M}_{h}}}^{*})]\hat{\psi }_{ex}^{\dagger }(-{{{\vec{k}}}_{||}},{{M}_{h}},{{R}_{i}},\varepsilon )+ \nonumber\\
+{{P}^{*}}_{cv}(0){{T}^{*}}\left( {{R}_{i}},\varepsilon ,-{{{\vec{k}}}_{||}} \right)[{{C}_{\vec{k},-}}(\vec{\sigma }_{{\vec{k}}}^{+}\cdot \vec{\sigma }_{{{M}_{h}}}^{{}})+{{C}_{\vec{k},+}}(\vec{\sigma }_{{\vec{k}}}^{-}\cdot \vec{\sigma }_{{{M}_{h}}}^{{}})]\hat{\psi }_{ex}^{{}}(-{{{\vec{k}}}_{||}},{{M}_{h}},{{R}_{i}},\varepsilon )\}
\end{eqnarray}
\end{widetext}
This expression is similar with the Hamiltonian in the absence of the RSOC. Now the Coulomb interaction between charged carriers in the presence of the RSOC will be investigated.
\section{The Coulomb interaction in the 2D electron-hole system under the influence of the Rashba spin-orbit coupling}
The Coulomb interaction in the 2D e-h system taking into account the Rashba spin-orbit coupling was discussed in Ref. [19, 22]. Below we will remember these results including all valence electron states (12). In the present description the multi-component electron field contains a larger variety of the valence band states in comparison with [19, 22]. For the very beginning the properties of the density operator of the electron field $\hat{\rho }(\vec{r})$ and of its Fourier components $\hat{\rho }(\vec{Q})$ will be discussed. To this end the Fermi-type creation and annihilation operators of the electron on different states were introduced. They are denoted as $a_{{{R}_{i}},g}^{\dagger },a_{{{R}_{i}},g}^{{}}$ for the conduction band Rashba-type states (16) $\left| {{\psi }_{c}}({{R}_{i}},g;r) \right\rangle $, as $a_{{{M}_{v}},{{\varepsilon }_{m}},g}^{\dagger },a_{{{M}_{v}},{{\varepsilon }_{m}},g}^{{}}$ for the spinor valence band states (12) $\left| {{\psi }_{v}}({{M}_{v}},{{\varepsilon }_{m}},g;r) \right\rangle $ and as $a_{{{M}_{v}},\varepsilon _{m}^{-},g}^{\dagger },a_{{{M}_{v}},\varepsilon _{m}^{-},g}^{{}}$ for another spinor valence band states (12) $\left| {{\psi }_{v}}({{M}_{v}},\varepsilon _{m}^{-},g;r) \right\rangle $. These spinor-type functions have a form of a column with two components corresponding to two spin projections on the direction of the magnetic field. The conjugate functions $\left\langle  {{\psi }_{c}}({{R}_{i}},g;r) \right|$, $\left\langle  {{\psi }_{c}}({{R}_{i}},g;r) \right|$ and $\left\langle  {{\psi }_{v}}({{M}_{v}},\varepsilon _{m}^{-},g;r) \right|$ have a form of a row with two components conjugate to the components of the columns. With the aid of the electron creation and annihilation operators and of the spinor-type wave functions the creation and annihilation operators ${{\hat{\Psi }}^{\dagger }}(r)$ and $\hat{\Psi }(r)$ of the multi-component electron field can be written as
\begin{eqnarray}
\hat{\Psi }(r)=\sum\limits_{i=1,2}{\sum\limits_{g}{{}}}\left| {{\psi }_{c}}({{R}_{i}},g;r) \right\rangle {{a}_{{{R}_{i}},g}}+ \nonumber \\
+\sum\limits_{{{M}_{v}}}{\sum\limits_{{{\varepsilon }_{m}}}{\sum\limits_{g}{{}}}}\left| {{\psi }_{v}}({{M}_{v}},{{\varepsilon }_{m}},g;r) \right\rangle {{a}_{{{M}_{v}},{{\varepsilon }_{m}},g}}+ \nonumber\\
+\sum\limits_{{{M}_{v}}}{\sum\limits_{\varepsilon _{m}^{-}}{\sum\limits_{g}{{}}}}\left| {{\psi }_{v}}({{M}_{v}},\varepsilon _{m}^{-},g;r) \right\rangle {{a}_{{{M}_{v}},\varepsilon _{m}^{-},g}}, \nonumber\\
{{{\hat{\Psi }}}^{\dagger }}(r)=\sum\limits_{i=1,2}{\sum\limits_{g}{{}}}\left\langle  {{\psi }_{c}}({{R}_{i}},g;r) \right|a_{{{R}_{i}},g}^{\dagger }+ \nonumber\\
+\sum\limits_{{{M}_{v}}}{\sum\limits_{{{\varepsilon }_{m}}}{\sum\limits_{g}{{}}}}\left\langle  {{\psi }_{v}}({{M}_{v}},{{\varepsilon }_{m}},g;r) \right|a_{{{M}_{v}},{{\varepsilon }_{m}},g}^{\dagger }+ \nonumber\\
+\sum\limits_{{{M}_{v}}}{\sum\limits_{\varepsilon _{m}^{-}}{\sum\limits_{g}{{}}}}\left\langle  {{\psi }_{v}}({{M}_{v}},\varepsilon _{m}^{-},g;r) \right|a_{{{M}_{v}},\varepsilon _{m}^{-},g}^{\dagger }
\end{eqnarray}
The density operator of the electron field $\hat{\rho }(\vec{r})$ and its Fourier components $\hat{\rho }(\vec{Q})$ are determined by the expressions
\begin{eqnarray}
\hat{\rho }(\vec{r})={{{\hat{\Psi }}}^{\dagger }}(r)\hat{\Psi }(r), \nonumber \\
\hat{\rho }(\vec{Q})=\int{{{d}^{2}}\vec{r}}\hat{\rho }(\vec{r}){{e}^{i\vec{Q}\vec{r}}}
\end{eqnarray}
The density operator looks as
\begin{widetext}
\begin{eqnarray}
\hat{\rho }(\vec{r})=\sum\limits_{i,j=1,2}{\sum\limits_{g,q}{{}}}a_{{{R}_{j}},q}^{\dagger }{{a}_{{{R}_{i}},g}}\left\langle {{\psi }_{c}}({{R}_{j}},q;r)|{{\psi }_{c}}({{R}_{i}},g;r) \right\rangle + \nonumber\\
+\sum\limits_{{{M}_{v}},{{{{M}'}}_{v}}}{\sum\limits_{{{\varepsilon }_{m}},{{\varepsilon }_{{{m}'}}}}{\sum\limits_{g,q}{{}}}}a_{{{{{M}'}}_{v}},{{\varepsilon }_{{{m}'}}},q}^{\dagger }{{a}_{{{M}_{v}},{{\varepsilon }_{m}},g}}\left\langle {{\psi }_{v}}({{{{M}'}}_{v}},{{\varepsilon }_{{{m}'}}},q;r) \right.\left| {{\psi }_{v}}({{M}_{v}},{{\varepsilon }_{m}},g;r) \right\rangle + \nonumber\\
+\sum\limits_{{{M}_{v}},{{{{M}'}}_{v}}}{\sum\limits_{\varepsilon _{m}^{-},\varepsilon _{{{m}'}}^{-}}{\sum\limits_{g,q}{{}}a_{{{{{M}'}}_{v}},\varepsilon _{{{m}'}}^{-},q}^{\dagger }{{a}_{{{M}_{v}},\varepsilon _{m}^{-},g}}}}\left\langle {{\psi }_{v}}({{{{M}'}}_{v}},\varepsilon _{{{m}'}}^{-},q;r) \right.\left| {{\psi }_{v}}({{M}_{v}},\varepsilon _{m}^{-},g;r) \right\rangle + \nonumber\\
+\sum\limits_{{{M}_{v}},{{{{M}'}}_{v}}}{\sum\limits_{\varepsilon _{{{m}'}}^{{}},\varepsilon _{m}^{-}}{\sum\limits_{g,q}{{}}a_{{{{{M}'}}_{v}},\varepsilon _{{{m}'}}^{{}},q}^{\dagger }{{a}_{{{M}_{v}},\varepsilon _{m}^{-},g}}}}\left\langle {{\psi }_{v}}({{{{M}'}}_{v}},\varepsilon _{{{m}'}}^{{}},q;r) \right.\left| {{\psi }_{v}}({{M}_{v}},\varepsilon _{m}^{-},g;r) \right\rangle + \nonumber\\
+\sum\limits_{{{M}_{v}},{{{{M}'}}_{v}}}{\sum\limits_{\varepsilon _{{{m}'}}^{-},\varepsilon _{m}^{{}}}{\sum\limits_{g,q}{{}}a_{{{{{M}'}}_{v}},\varepsilon _{{{m}'}}^{-},q}^{\dagger }{{a}_{{{M}_{v}},\varepsilon _{m}^{{}},g}}}}\left\langle {{\psi }_{v}}({{{{M}'}}_{v}},\varepsilon _{{{m}'}}^{-},q;r) \right.\left| {{\psi }_{v}}({{M}_{v}},\varepsilon _{m}^{{}},g;r) \right\rangle + \nonumber\\
+\sum\limits_{i=1,2}{\sum\limits_{{{M}_{v}}}{\sum\limits_{{{\varepsilon }_{m}}}{\sum\limits_{g,q}{{}}}}}a_{{{R}_{i}},q}^{\dagger }{{a}_{{{M}_{v}},{{\varepsilon }_{m}},g}}\left\langle {{\psi }_{c}}({{R}_{i}},q;r)|{{\psi }_{v}}({{M}_{v}},{{\varepsilon }_{m}},g;r) \right\rangle + \nonumber\\
+\sum\limits_{i=1,2}{\sum\limits_{{{M}_{v}}}{\sum\limits_{\varepsilon _{m}^{-}}{\sum\limits_{g,q}{{}}}}}a_{{{R}_{i}},q}^{\dagger }{{a}_{{{M}_{v}},\varepsilon _{m}^{-},g}}\left\langle {{\psi }_{c}}({{R}_{i}},q;r)|{{\psi }_{v}}({{M}_{v}},\varepsilon _{m}^{-},g;r) \right\rangle + \nonumber\\
+\sum\limits_{i=1,2}{\sum\limits_{{{M}_{v}}}{\sum\limits_{{{\varepsilon }_{m}}}{\sum\limits_{g,q}{{}}}}}a_{{{M}_{v}},{{\varepsilon }_{m}},q}^{\dagger }{{a}_{{{R}_{i}},g}}\left\langle {{\psi }_{v}}({{M}_{v}},{{\varepsilon }_{m}},q;r)|{{\psi }_{c}}({{R}_{i}},g;r) \right\rangle + \nonumber\\
+\sum\limits_{i=1,2}{\sum\limits_{{{M}_{v}}}{\sum\limits_{\varepsilon _{m}^{-}}{\sum\limits_{g,q}{{}}}}}a_{{{M}_{v}},\varepsilon _{m}^{-},q}^{\dagger }{{a}_{{{R}_{i}},g}}\left\langle {{\psi }_{v}}({{M}_{v}},\varepsilon _{m}^{-},q;r)|{{\psi }_{c}}({{R}_{i}},g;r) \right\rangle
\end{eqnarray}
\end{widetext}
The Fourier components $\hat{\rho }(\vec{Q})$ of the density operator determine the Coulomb interaction between the electrons. They will be calculated below taking into account the spinor-type wave functions (12) and (16). For example, the first term in the expressions (38) looks as
\begin{eqnarray}
&{{{\hat{\rho }}}_{c-c}}({{R}_{1}};{{R}_{1}};\vec{Q})=& \nonumber \\
& =\sum\limits_{q,g}{a_{{{R}_{1}},q}^{\dagger }{{a}_{{{R}_{1}},g}}}\left\langle  {{\psi }_{c}}({{R}_{1}},q;\vec{r}) \right|\left. {{\psi }_{c}}({{R}_{1}},g;\vec{r}) \right\rangle =& \nonumber \\
& =\sum\limits_{q,g}{a_{{{R}_{1}},q}^{\dagger }{{a}_{{{R}_{1}},g}}}\int{{{d}^{2}}\vec{r}}U_{c,s,q}^{*}(\vec{r}){{U}_{c,s,g}}(\vec{r})\frac{{{e}^{i(g+{{Q}_{x}}-q)x}}}{{{L}_{x}}}\times & \nonumber \\
& \times [{{\left| {{a}_{0}} \right|}^{2}}\varphi _{0}^{*}(y-q l_{0}^{2})\varphi _{0}^{{}}(y-g l_{0}^{2})+ & \nonumber \\
& +{{\left| {{b}_{1}} \right|}^{2}}\varphi _{1}^{*}(y-q l_{0}^{2})\varphi _{1}^{{}}(y-gl_{0}^{2})], & \nonumber \\
 & \vec{r}=\vec{R}+\vec{\rho }&
\end{eqnarray}
Following the formula (22) it is necessary to separate the integration of the quickly varying periodic parts on the volume ${{v}_{0}}$ of the elementary lattice cell and the integration of the slowly varying envelope parts on the lattice point vectors $\vec{R}$ as follows
\begin{eqnarray}
&\frac{1}{{{v}_{0}}}\int\limits_{{{v}_{0}}}{d\rho }U_{c,s,g+{{Q}_{x}}}^{*}(\rho )U_{c,s,g}^{{}}(\rho ){{e}^{i{{Q}_{y}}{{\rho }_{y}}}}=1+O(\vec{Q}),& \nonumber \\
& \frac{1}{{{L}_{x}}}\int{{{e}^{i(g-q+{{Q}_{x}}){{R}_{x}}}}d{{R}_{x}}}={{\delta }_{kr}}(q,g+{{Q}_{x}}),& \nonumber \\
& \tilde{\phi }(n,m;\vec{Q})= & \nonumber\\
& =\int{d{{R}_{y}}}\varphi _{n}^{*}\left( {{R}_{y}}-\frac{{{Q}_{x}}l_{0}^{2}}{2} \right)\varphi _{m}^{{}}\left( {{R}_{y}}+\frac{{{Q}_{x}}l_{0}^{2}}{2} \right){{e}^{i{{Q}_{y}}{{R}_{y}}}}= & \nonumber \\
 & ={{{\tilde{\phi }}}^{*}}(m,n;-\vec{Q})&
\end{eqnarray}
Here $O(\vec{Q})$ is an infinitesimal value much smaller than unity, tending to zero in the limit $Q\to 0$. It will be neglected in all calculations below. The calculation give rise to the final form
\begin{eqnarray}
&{{{\hat{\rho }}}_{c-c}}({{R}_{1}};{{R}_{1}};\vec{Q})=[{{\left| {{a}_{0}} \right|}^{2}}\tilde{\phi }(0,0;\vec{Q})+& \nonumber \\
 & +{{\left| {{b}_{1}} \right|}^{2}}\tilde{\phi }(1,1;\vec{Q})]\hat{\rho }({{R}_{1}};{{R}_{1}};\vec{Q})= & \nonumber \\
 & =\tilde{S}({{R}_{1}};{{R}_{1}};\vec{Q})\hat{\rho }({{R}_{1}};{{R}_{1}};\vec{Q}),&  \nonumber \\
 & \hat{\rho }({{R}_{1}};{{R}_{1}};\vec{Q})=\sum\limits_{t}{{}}{{e}^{i{{Q}_{y}}tl_{0}^{2}}}a_{{{R}_{1}},t+\frac{{{Q}_{x}}}{2}}^{\dagger }a_{{{R}_{1}},t-\frac{{{Q}_{x}}}{2}}^{{}}, & \nonumber \\
 & \tilde{S}({{R}_{1}};{{R}_{1}};\vec{Q})=[{{\left| {{a}_{0}} \right|}^{2}}\tilde{\phi }(0,0;\vec{Q})+{{\left| {{b}_{1}} \right|}^{2}}\tilde{\phi }(1,1;\vec{Q})]&
\end{eqnarray}
The expression (41) looks as a product of one numeral factor $\tilde{S}({{R}_{1}};{{R}_{1}};\vec{Q})$, which concerns the concrete electron spinor state and another operator type factor of the general form
\begin{eqnarray}
\hat{\rho }(\xi ,\eta ;\vec{Q})= \nonumber \\
=\sum\limits_{t}{{}}{{e}^{i{{Q}_{y}}t l_{0}^{2}}}a_{\xi ,t+\frac{{{Q}_{x}}}{2}}^{\dagger }{{a}_{\eta ,t-\frac{{{Q}_{x}}}{2}}}= \nonumber \\
={{{\hat{\rho }}}^{\dagger }}(\eta ,\xi ;-\vec{Q})
\end{eqnarray}
It will be met in all expressions listed below, but with different meanings of $\xi $ and $\eta $, as follows
\begin{eqnarray}
&{{{\hat{\rho }}}_{c-c}}({{R}_{2}};{{R}_{2}};\vec{Q})=\tilde{S}({{R}_{2}};{{R}_{2}};\vec{Q})\hat{\rho }({{R}_{2}};{{R}_{2}};\vec{Q}),& \nonumber \\
& \tilde{S}({{R}_{2}};{{R}_{2}};\vec{Q})=\tilde{\phi }(0,0;\vec{Q}),& \nonumber \\
& {{{\hat{\rho }}}_{c-c}}({{R}_{1}};{{R}_{2}};\vec{Q})=\tilde{S}({{R}_{1}};{{R}_{2}};\vec{Q})\hat{\rho }({{R}_{1}};{{R}_{2}};\vec{Q}),& \nonumber \\
& \tilde{S}({{R}_{1}};{{R}_{2}};\vec{Q})=b_{1}^{*}\tilde{\phi }(1,0;\vec{Q}),& \nonumber \\
& {{{\hat{\rho }}}_{c-c}}({{R}_{2}};{{R}_{1}};\vec{Q})=\tilde{S}({{R}_{2}};{{R}_{1}};\vec{Q})\hat{\rho }({{R}_{2}};{{R}_{1}};\vec{Q})& \nonumber \\
& =\rho _{c-c}^{\dagger }({{R}_{1}};{{R}_{2}};-\vec{Q}),& \nonumber \\
& \tilde{S}({{R}_{2}};{{R}_{1}};\vec{Q})=b_{1}^{{}}\tilde{\phi }(0,1;\vec{Q}) &
\end{eqnarray}
One of the valence electron density fluctuation operator looks as
\begin{eqnarray}
&{{{\hat{\rho }}}_{v-v}}({{M}_{v}},\varepsilon _{m}^{-},{{{{M}'}}_{v}},\varepsilon _{{{m}'}}^{-};\vec{Q})= & \nonumber \\
 & \sum\limits_{q,g}{{}}a_{{{M}_{v}},\varepsilon _{m}^{-},q}^{\dagger }{{a}_{{{{{M}'}}_{v}},\varepsilon _{{{m}'}}^{-},g}}\int{{{d}^{2}}\vec{r}}{{e}^{i\vec{Q}\vec{r}}}\times & \nonumber \\
 & \times \left\langle  {{\psi }_{v}}({{M}_{v}},\varepsilon _{m}^{-},q;\vec{r}) \right|\left. {{\psi }_{v}}({{{{M}'}}_{v}},\varepsilon _{{{m}'}}^{-},g;\vec{r}) \right\rangle = & \nonumber \\
 & =\tilde{S}({{M}_{v}},\varepsilon _{m}^{-};{{{{M}'}}_{v}},\varepsilon _{{{m}'}}^{-};\vec{Q})\hat{\rho }({{M}_{v}},\varepsilon _{m}^{-};{{M}_{v}},\varepsilon _{{{m}'}}^{-}\vec{Q}), & \nonumber \\
 & \tilde{S}({{M}_{v}},\varepsilon _{m}^{-};{{{{M}'}}_{v}},\varepsilon _{{{m}'}}^{-};\vec{Q})= & \nonumber \\
 & ={{\delta }_{{{M}_{v}},{{{{M}'}}_{v}}}}[d_{m-3}^{-}d_{{m}'-3}^{-*}\tilde{\phi }(m-3,{m}'-3;\vec{Q})+ & \nonumber \\
 & +c_{m}^{-}c_{{{m}'}}^{-*}\tilde{\phi }(m,{m}';\vec{Q})], & \nonumber \\
 & m,{m}'\ge 3&
\end{eqnarray}
Here we have taken into account the following property of the valence band periodic parts
\begin{eqnarray}
&\frac{1}{{{v}_{0}}}\int\limits_{{{v}_{0}}}{d\vec{\rho }}U_{v,p,i,g+{{Q}_{x}}}^{*}(\rho )U_{v,p,j,g}^{{}}(\rho ){{e}^{i{{Q}_{y}}{{\rho }_{y}}}}=& \nonumber \\
 & ={{\delta }_{ij}}+O(\vec{Q}),& \nonumber \\
 & i,j=x,y&
\end{eqnarray}
They leads to the Kronekker symbol ${{\delta }_{{{M}_{v}},{{{{M}'}}_{v}}}}$ in the expression (44) and in the next ones concerning the valence band, as follows
\begin{eqnarray}
&{{{\hat{\rho }}}_{v-v}}({{M}_{v}},\varepsilon _{m}^{{}};{{{{M}'}}_{v}},\varepsilon _{{{m}'}}^{{}};\vec{Q})=& \nonumber \\
 & =\tilde{S}({{M}_{v}},\varepsilon _{m}^{{}};{{{{M}'}}_{v}},\varepsilon _{{{m}'}}^{{}};\vec{Q})\hat{\rho }({{M}_{v}},\varepsilon _{m}^{{}};{{M}_{v}},\varepsilon _{{{m}'}}^{{}};\vec{Q}),& \nonumber \\
 & \tilde{S}({{M}_{v}},\varepsilon _{m}^{{}};{{{{M}'}}_{v}},\varepsilon _{{{m}'}}^{{}};\vec{Q})={{\delta }_{{{M}_{v}},{{{{M}'}}_{v}}}}\tilde{\phi }(m,{m}';\vec{Q}), & \nonumber \\
 & m,{m}'=0,1,2,& \nonumber \\
 & {{{\hat{\rho }}}_{v-v}}({{{{M}'}}_{v}},\varepsilon _{{{m}'}}^{{}};{{M}_{v}},\varepsilon _{m}^{-};\vec{Q})= & \nonumber \\
 & =\tilde{S}({{{{M}'}}_{v}},\varepsilon _{{{m}'}}^{{}};{{M}_{v}},\varepsilon _{m}^{-};\vec{Q})\hat{\rho }({{M}_{v}},\varepsilon _{{{m}'}}^{{}};{{M}_{v}},\varepsilon _{m}^{-};\vec{Q}), & \nonumber \\
 & \tilde{S}({{{{M}'}}_{v}},\varepsilon _{{{m}'}}^{{}};{{M}_{v}},\varepsilon _{m}^{-};\vec{Q})={{\delta }_{{{M}_{v}},{{{{M}'}}_{v}}}}c_{m}^{-*}\tilde{\phi }({m}',m;\vec{Q}),& \nonumber \\
 & {m}'=0,1,2,\text{ }m\ge 3,& \nonumber \\
 & {{{\hat{\rho }}}_{v-v}}({{M}_{v}},\varepsilon _{m}^{-};{{{{M}'}}_{v}},\varepsilon _{{{m}'}}^{{}};\vec{Q})= & \nonumber\\
 & =\tilde{S}({{M}_{v}},\varepsilon _{m}^{-};{{{{M}'}}_{v}},\varepsilon _{{{m}'}}^{{}};\vec{Q})\hat{\rho }({{M}_{v}},\varepsilon _{m}^{-};{{M}_{v}},\varepsilon _{{{m}'}}^{{}};\vec{Q}), & \nonumber\\
 & \tilde{S}({{M}_{v}},\varepsilon _{m}^{-};{{{{M}'}}_{v}},\varepsilon _{{{m}'}}^{{}};\vec{Q})={{\delta }_{{{M}_{v}},{{{{M}'}}_{v}}}}c_{m}^{-}\tilde{\phi }(m,{m}';\vec{Q}), & \nonumber\\
 & m\ge 3,\text{ }{m}'=0,1,2&
\end{eqnarray}
As usual they obey to the equalities
\begin{eqnarray}
\hat{\rho }_{v-v}^{\dagger }(\xi ;\eta ;\vec{Q})=\hat{\rho }_{v-v}^{{}}(\eta ;\xi ;-\vec{Q}), \nonumber\\
\hat{\rho }(\xi ;\eta ;\vec{Q})=\hat{\rho }_{{}}^{\dagger }(\eta ;\xi ;-\vec{Q}), \nonumber \\
\tilde{\phi }(n;m;\vec{Q})={{{\tilde{\phi }}}^{*}}(m;n;-\vec{Q}), \nonumber \\
\tilde{S}(\xi ;\eta ;\vec{Q})={{{\tilde{S}}}^{*}}(\eta ;\xi ;-\vec{Q})
\end{eqnarray}
Up till now we have deal with the intraband density operators ${{\hat{\rho }}_{c-c}}(\xi ;\eta ;\vec{Q})$ and ${{\hat{\rho }}_{v-v}}(\xi ;\eta ;\vec{Q})$. The interband density operators ${{\hat{\rho }}_{c-v}}(\xi ;\eta ;\vec{Q})$ and ${{\hat{\rho }}_{v-c}}(\xi ;\eta ;\vec{Q})$ depend on the interband exchange electron densities of the type $U_{c,s,g+{{Q}_{x}}}^{*}(\rho )\frac{1}{\sqrt{2}}({{U}_{v,p,x,g}}(\rho )\pm i{{U}_{v,p,y,g}}(\rho ))$ and its complex conjugate value. They contain the quickly oscillating periodic parts with different parities and the ortoganality integral on the elementary lattice cell has an infinitesimal value
\begin{eqnarray}
O(\vec{Q})=\frac{1}{{{v}_{0}}}\int\limits_{{{v}_{0}}}{d\rho }U_{c,s,g+{{Q}_{x}}}^{*}(\rho ){{e}^{i{{Q}_{y}}{{\rho }_{y}}}}\times  \nonumber \\
\times \frac{1}{\sqrt{2}}({{U}_{v,p,x,g}}(\rho )\pm i{{U}_{v,p,y,g}}(\rho ))
\end{eqnarray}
 This integral is different from zero if one takes into account, for example, the term $i{{Q}_{y}}{{\rho }_{y}}$ appearing in the series expansion of the function ${{e}^{i{{Q}_{y}}{{\rho }_{y}}}}$. It gives rise to the interband dipole momentum ${{\vec{d}}_{cv}}$ with the component
 \begin{eqnarray}
 {{d}_{cv,y}}=\frac{e}{{{v}_{0}}}\int\limits_{{{v}_{0}}}{d\rho }U_{c,s,g}^{*}(\rho ){{\rho }_{y}}{{U}_{v,p,y,g}}(\rho ), \nonumber \\
  O(\vec{Q})\approx {{Q}_{y}}{{d}_{cv,y}}
 \end{eqnarray}
The Coulomb interaction depending on the interband exchange electron densities ${{\hat{\rho }}_{cv}}(\vec{Q}){{\hat{\rho }}_{vc}}(-\vec{Q})$ has a form of the dipole-dipole interaction instead of the charge-charge interaction, which takes place only in the intraband cases. It is known as a long-range Coulomb interaction and gives rise to the longitudinal-transverse splitting of the three-fold degenerate levels of the dipole-active excitons in the cubic crystals [27]. Such type effects with the participation of the 2D magnetoexcitons were not investigated up till now, to the best of our knowledge, and remain outside the present review article.

The density operator $\hat{\rho }(\vec{Q})$ in the frame of the electron spinor states (12) and (16) looks as
\begin{eqnarray}
\hat{\rho }(\vec{Q})=\sum\limits_{i,j}{{}}{{{\hat{\rho }}}_{c-c}}({{R}_{i}};{{R}_{j}};\vec{Q})+ \nonumber\\
+\sum\limits_{{{M}_{v}},{{{{M}'}}_{v}}}{\sum\limits_{{{\varepsilon }_{m}},{{\varepsilon }_{{{m}'}}}}{{}}}{{{\hat{\rho }}}_{v-v}}({{M}_{v}},{{\varepsilon }_{m}};{{{{M}'}}_{v}},{{\varepsilon }_{{{m}'}}};\vec{Q})+ \nonumber\\
+\sum\limits_{{{M}_{v}},{{{{M}'}}_{v}}}{\sum\limits_{\varepsilon _{m}^{-},\varepsilon _{{{m}'}}^{-}}{{}}}{{{\hat{\rho }}}_{v-v}}({{M}_{v}},\varepsilon _{m}^{-};{{{{M}'}}_{v}},\varepsilon _{{{m}'}}^{-};\vec{Q})+ \nonumber\\
+\sum\limits_{{{M}_{v}},{{{{M}'}}_{v}}}{\sum\limits_{\varepsilon _{m}^{-},\varepsilon _{{{m}'}}^{{}}}{{}}}{{{\hat{\rho }}}_{v-v}}({{{{M}'}}_{v}},\varepsilon _{{{m}'}}^{{}};{{M}_{v}},\varepsilon _{m}^{-};\vec{Q})+ \nonumber\\
+\sum\limits_{{{M}_{v}},{{{{M}'}}_{v}}}{\sum\limits_{\varepsilon _{m}^{-},\varepsilon _{{{m}'}}^{{}}}{{}}}{{{\hat{\rho }}}_{v-v}}({{M}_{v}},\varepsilon _{m}^{-};{{{{M}'}}_{v}},\varepsilon _{{{m}'}}^{{}};\vec{Q})+ \nonumber\\
+\sum\limits_{i}{\sum\limits_{{{M}_{v}}}{\sum\limits_{{{\varepsilon }_{m}}}{{}}}}{{{\hat{\rho }}}_{c-v}}({{R}_{i}};{{M}_{v}},{{\varepsilon }_{m}};\vec{Q})+ \nonumber\\
+\sum\limits_{i}{\sum\limits_{{{M}_{v}}}{\sum\limits_{\varepsilon _{m}^{-}}{{}}}}{{{\hat{\rho }}}_{c-v}}({{R}_{i}};{{M}_{v}},\varepsilon _{m}^{-};\vec{Q})+ \nonumber\\
+\sum\limits_{i}{\sum\limits_{{{M}_{v}}}{\sum\limits_{{{\varepsilon }_{m}}}{{}}}}{{{\hat{\rho }}}_{v-c}}({{M}_{v}},{{\varepsilon }_{m}};{{R}_{i}};\vec{Q})+ \nonumber\\
+\sum\limits_{i}{\sum\limits_{{{M}_{v}}}{\sum\limits_{\varepsilon _{m}^{-}}{{}}}}{{{\hat{\rho }}}_{v-c}}({{M}_{v}},\varepsilon _{m}^{-};{{R}_{i}};\vec{Q})
\end{eqnarray}
The first five terms of this expression depend on the intraband electron densities and determine the charge-charge Coulomb interaction. The last four terms depend on the interband electron densities and lead to the dipole-dipole long-range Coulomb interaction.

The strength of the Coulomb interaction is determined by coefficients ${{a}_{n}},{{b}_{n}},{{c}_{n}},{{d}_{n}}$ of the spinor-type wave function (12) and (16) as well as by the normalization and orthogonality-type integrals $\tilde{\phi }(n,m,\vec{Q})$. They have the properties:
\begin{eqnarray}
\tilde{\phi }(n,m,\vec{Q})={{e}^{-\frac{{{Q}^{2}}{{l}^{2}}}{4}}}{{A}_{n,m}}(\vec{Q}) \nonumber\\
{{A}_{n,m}}(0)={{\delta }_{n,m}}
\end{eqnarray}
The diagonal coefficients ${{A}_{n,n}}(\vec{Q})$ with $n=0,1,3$ will be calculated below. The nondiagonal coefficients with $n\ne m$ in the limit $Q\to 0$ are proportional to the vector components ${{Q}_{i}}$ in a degree $|n-m|$. They can be neglected in the zeroth order approximation together with another corrections denoted as $O(\vec{Q})$. It essentially deminish the number of the actual components of the density operator $\hat{\rho }(\vec{Q})$.

In the zeroth order approximation, neglecting the corrections of the order $O(\vec{Q})$ we will deal only with diagonal terms which permit the simplified denotations
\begin{eqnarray}
\hat{\rho }(\xi ;\xi ;\vec{Q})=\hat{\rho }(\xi ;\vec{Q}), \nonumber \\
\tilde{S}(\xi ;\xi ;\vec{Q})=\tilde{S}(\xi ;\vec{Q})={{e}^{-\frac{{{Q}^{2}}l_{0}^{2}}{4}}}S(\xi ;\vec{Q})
\end{eqnarray}
The concrete values of the coefficients $S(\xi ;\vec{Q})$ are
\begin{eqnarray}
& S({{R}_{1}};\vec{Q})=[|{{a}_{0}}{{|}^{2}}{{A}_{0,0}}(\vec{Q})+|{{b}_{1}}{{|}^{2}}{{A}_{1,1}}(\vec{Q})],& \nonumber \\
& S({{R}_{2}};\vec{Q})={{A}_{0,0}}(\vec{Q}), & \nonumber\\
& S({{\varepsilon }_{m}};\vec{Q})={{A}_{m,m}}(\vec{Q}),m=0,1,2,& \nonumber \\
& S(\varepsilon _{m}^{-};\vec{Q})= & \nonumber\\
& =[|d_{m-3}^{-}{{|}^{2}}{{A}_{m-3,m-3}}(\vec{Q})+|c_{m}^{-}{{|}^{2}}{{A}_{m,m}}(\vec{Q})],& \nonumber \\
& m\ge 3&
\end{eqnarray}
The calculated values ${{A}_{m,m}}(\vec{Q})$ equal to
\begin{eqnarray}
& {{A}_{0,0}}(\vec{Q})=1,\text{ }{{A}_{1,1}}(\vec{Q})=\left( 1-\frac{{{Q}^{2}}l_{0}^{2}}{2} \right), & \nonumber \\
& {{A}_{3,3}}(\vec{Q})=1-\frac{3}{2}{{Q}^{2}}l_{0}^{2}+\frac{3}{8}{{Q}^{4}}l_{0}^{4}-\frac{1}{48}{{Q}^{6}}l_{0}^{6} &
\end{eqnarray}
The diagonal part of the density operator $\hat{\rho }(\vec{Q})$ looks as
\begin{eqnarray}
& \hat{\rho }(\vec{Q})={{e}^{-\frac{{{Q}^{2}}l_{0}^{2}}{4}}}\{\sum\limits_{i}{{}}S({{R}_{i}};\vec{Q})\hat{\rho }({{R}_{i}};\vec{Q})+ & \nonumber \\
& +\sum\limits_{{{M}_{v}}}{\sum\limits_{{{\varepsilon }_{m}}}{{}}S({{M}_{v}},{{\varepsilon }_{m}};\vec{Q})}\hat{\rho }({{M}_{v}},{{\varepsilon }_{m}};\vec{Q})+ & \nonumber \\
& +\sum\limits_{{{M}_{v}}}{\sum\limits_{\varepsilon _{m}^{-}}{{}}S({{M}_{v}},\varepsilon _{m}^{-};\vec{Q})}\hat{\rho }({{M}_{v}},\varepsilon _{m}^{-};\vec{Q})\}&
\end{eqnarray}
It contains two separate contributions from the conduction and valence bands. The latter contribution in its turn can be represented as due to the electrons of the full filled valence band extracting the contribution of the holes created in its frame. To show it one can introduce the hole creation and annihilation operators as follows
\begin{eqnarray}
b_{{{M}_{h}},\varepsilon ,q}^{\dagger }={{a}_{v,-{{M}_{v}},\varepsilon ,-q}}, \nonumber \\
{{b}_{{{M}_{h}},\varepsilon ,q}}=a_{v,-{{M}_{v}},\varepsilon ,-q}^{\dagger }, \nonumber \\
\varepsilon ={{\varepsilon }_{m}},m=0,1,2, \nonumber \\
\varepsilon =\varepsilon _{m}^{-},m\ge 3
\end{eqnarray}
It leads to the relation
\begin{eqnarray}
& {{{\hat{\rho }}}_{v}}(-{{M}_{v}},\varepsilon ,\vec{Q})=N{{\delta }_{kr}}(\vec{Q},0)-{{{\hat{\rho }}}_{h}}({{M}_{h}},\varepsilon ,\vec{Q}), & \nonumber \\
& N=\frac{S}{2\pi l_{0}^{2}}&
\end{eqnarray}
where the hole density operator ${{\rho }_{h}}({{M}_{h}},\varepsilon ,\vec{Q})$ looks as
\begin{equation}
{{\hat{\rho }}_{h}}({{M}_{h}},\varepsilon ,\vec{Q})=\sum\limits_{t}{{e}^{-i{{Q}_{y}}tl_{0}^{2}}}b_{{{M}_{h}},\varepsilon ,t+\frac{{{Q}_{x}}}{2}}^{\dagger }b_{{{M}_{h}},\varepsilon ,t-\frac{{{Q}_{x}}}{2}}
\end{equation}
The constant part $N{{\delta }_{kr}}(\vec{Q},0)$in (57) created by electron of the full filled valence band is compensated by the influence of the positive electric charges of the background nuclei. In the jelly model of the system their presence is taken into account excluding from the Hamiltonian of the Coulomb interaction the point $\vec{Q}=0$ [29]. Taking into account the fully neutral system of the bare electrons and of the positive jelly background we will operate only with the conduction band electrons and with the holes in the valence band. In this electron-hole description the density operator $\hat{\rho }(\vec{Q})$ becomes equal
\begin{eqnarray}
\hat{\rho }(\vec{Q})={{{\hat{\rho }}}_{e}}(\vec{Q})-{{{\hat{\rho }}}_{h}}(\vec{Q}), \nonumber \\
\vec{Q}\ne 0
\end{eqnarray}
where
\begin{eqnarray}
& {{{\hat{\rho }}}_{h}}(\vec{Q})=\sum\limits_{{{M}_{h}},{{\varepsilon }_{m}}}{{}}\tilde{S}({{\varepsilon }_{m}};\vec{Q}){{{\hat{\rho }}}_{h}}({{M}_{h}},{{\varepsilon }_{m}};\vec{Q})+& \nonumber \\
& +\sum\limits_{{{M}_{h}},\varepsilon _{m}^{-}}{{}}\tilde{S}(\varepsilon _{m}^{-};\vec{Q}){{{\hat{\rho }}}_{h}}({{M}_{h}},\varepsilon _{m}^{-};\vec{Q})= & \nonumber \\
& ={{e}^{-\frac{{{Q}^{2}}l_{0}^{2}}{4}}}\{\sum\limits_{{{M}_{h}},{{\varepsilon }_{m}}}{{}}S({{\varepsilon }_{m}};\vec{Q}){{{\hat{\rho }}}_{h}}({{M}_{h}},{{\varepsilon }_{m}};\vec{Q})+ & \nonumber \\
& +\sum\limits_{{{M}_{h}},\varepsilon _{m}^{-}}{{}}S(\varepsilon _{m}^{-};\vec{Q}){{{\hat{\rho }}}_{h}}({{M}_{h}},\varepsilon _{m}^{-};\vec{Q})\}, & \nonumber \\
& {{{\hat{\rho }}}_{e}}(\vec{Q})={{{\hat{\rho }}}_{c}}(\vec{Q})= & \nonumber \\
& =\sum\limits_{i=1,2}{{}}\tilde{S}({{R}_{i}};\vec{Q}){{{\hat{\rho }}}_{e}}({{R}_{i}};\vec{Q})= & \nonumber \\
& ={{e}^{-\frac{{{Q}^{2}}l_{0}^{2}}{4}}}\sum\limits_{i=1,2}{{}}S({{R}_{i}};\vec{Q}){{{\hat{\rho }}}_{e}}({{R}_{i}};\vec{Q}) &
\end{eqnarray}
The Hamiltonian of the Coulomb interaction of the initial bare electrons can be expressed through the electron field and density operators (36) and (50), as follows
\begin{eqnarray}
& {{H}_{Coul}}= & \nonumber \\
& =\frac{1}{2}\int{d\vec{1}}\int{d\vec{2}}{{{\hat{\Psi }}}^{\dagger }}(1){{{\hat{\Psi }}}^{\dagger }}(2)V(1-2)\hat{\Psi }(2)\hat{\Psi }(1)= & \nonumber \\
& =\frac{1}{2}\sum\limits_{{\vec{Q}}}{{}}V(\vec{Q})\int{d\vec{1}}\int{d\vec{2}}{{e}^{i\vec{Q}(\vec{1}-\vec{2})}}{{{\hat{\Psi }}}^{\dagger }}(1)\hat{\rho }(2)\hat{\Psi }(1)= & \nonumber \\
& =\frac{1}{2}\sum\limits_{{\vec{Q}}}{{}}V(\vec{Q})\int{\int{d\vec{r}}}{{e}^{i\vec{Q}\vec{r}}}{{{\hat{\Psi }}}^{\dagger }}(\vec{r})\hat{\rho }(-\vec{Q})\hat{\Psi }(\vec{r}), & \nonumber \\
& V(\vec{Q})=\frac{2\pi {{e}^{2}}}{{{\varepsilon }_{0}}S|\vec{Q}|}&
\end{eqnarray}
$V(\vec{Q})$ is the Fourier transform of the Coulomb interaction of the electrons situated on the surface of the 2D layer with the area S and dielectric constant ${{\varepsilon }_{0}}$ of the medium. The expression ${{\hat{\Psi }}^{\dagger }}(\vec{r})\hat{\rho }(-\vec{Q})\hat{\Psi }(\vec{r})$ contains the density operator $\hat{\rho }(-\vec{Q})$ intercalated between the field operators ${{\hat{\Psi }}^{\dagger }}(\vec{r})$ and $\hat{\Psi }(\vec{r})$. The operator $\hat{\Psi }(\vec{r})$ cannot be transposed over the operator $\hat{\rho }(-\vec{Q})$ because they do not commute, but its nonoperator part expressed through the spinor-type wave function can be transposed forming together with the conjugate wave function of the field operator ${{\hat{\Psi }}^{\dagger }}(\vec{r})$ a scalar.
After the integration on the coordinate $\vec{r}$ the quadratic intercalated density operators will appear with the forms
\begin{eqnarray}
& K(\xi ;\eta ;x;\eta ;\vec{Q})=& \\
& =\sum\limits_{t}{{}}{{e}^{i{{Q}_{y}}t l_{0}^{2}}}a_{\xi ,t+\frac{{{Q}_{x}}}{2}}^{\dagger }\hat{\rho }(x;y;-\vec{Q})a_{\eta ,t-\frac{{{Q}_{x}}}{2}}^{{}}= & \nonumber \\
& =\sum\limits_{t}{\sum\limits_{s}{{}}}{{e}^{i{{Q}_{\eta }}t l_{0}^{2}}}{{e}^{-i{{Q}_{y}}s l_{0}^{2}}}a_{\xi ,t+\frac{{{Q}_{x}}}{2}}^{\dagger }a_{x,s-\frac{{{Q}_{x}}}{2}}^{\dagger }a_{y,s+\frac{{{Q}_{x}}}{2}}^{{}}a_{\eta ,t-\frac{{{Q}_{x}}}{2}}^{{}}= & \nonumber \\
& =\hat{\rho }(\xi ;\eta ;\vec{Q})\hat{\rho }(x;y;-\vec{Q})-{{\delta }_{\eta ,x}}\hat{\rho }(\xi ;y;0) & \nonumber
\end{eqnarray}
The same relations remain in the electron-hole description.
The commutation relations between the density operators are the followings
\begin{eqnarray}
\hat{\rho }(\xi ;\eta ;\vec{Q})=\sum\limits_{t}{{e}^{i{{Q}_{y}}t l_{0}^{2}}}a_{\xi ,t+\frac{{{Q}_{x}}}{2}}^{\dagger }a_{\eta ,t-\frac{{{Q}_{x}}}{2}}, \nonumber\\
\hat{\rho }(x;y;\vec{P})=\sum\limits_{t}{{}}{{e}^{i{{P}_{y}}t l_{0}^{2}}}a_{x,t+\frac{{{P}_{x}}}{2}}^{\dagger }a_{y,t-\frac{{{P}_{x}}}{2}},\nonumber \\
\left[\hat{\rho }(\xi ;\eta ;\vec{Q}),\hat{\rho }(x;y;\vec{P})\right]= \nonumber \\
={{\delta }_{x,\eta }}\hat{\rho }(\xi ;y;\vec{P}+\vec{Q}){{e}^{\frac{i{{(\vec{P}\times \vec{Q})}_{z}}l_{0}^{2}}{2}}}- \nonumber \\
-{{\delta }_{\xi ,y}}\hat{\rho }(x;\eta ;\vec{P}+\vec{Q}){{e}^{\frac{-i{{(\vec{P}\times \vec{Q})}_{z}}l_{0}^{2}}{2}}}= \nonumber \\
=\cos \left( \frac{{{[\vec{P}\times \vec{Q}]}_{z}}l_{0}^{2}}{2} \right)[{{\delta }_{x,\eta }}\hat{\rho }(\xi ;y;\vec{P}+\vec{Q})- \nonumber \\
-{{\delta }_{\xi ,y}}\hat{\rho }(x;\eta ;\vec{P}+\vec{Q})]+ \nonumber \\
+i\sin \left( \frac{{{(\vec{P}\times \vec{Q})}_{z}}l_{0}^{2}}{2} \right)[{{\delta }_{x,\eta }}\hat{\rho }(\xi ;y;\vec{P}+\vec{Q})+ \nonumber \\
+{{\delta }_{\xi ,y}}\hat{\rho }(x;\eta ;\vec{P}+\vec{Q})]
\end{eqnarray}
The factor ${{e}^{-\frac{{{Q}^{2}}l_{0}^{2}}{4}}}$ arising from the product of the density operators $\hat{\rho }(\vec{Q})$ and $\hat{\rho }(-\vec{Q})$ being multiplied with the coefficient $V(\vec{Q})$ gives rise to the coefficient $W(\vec{Q})$ describing the effective Coulomb interaction in the conditions of the Landau quantization
\begin{equation}
W(\vec{Q})=V(\vec{Q}){{e}^{-\frac{{{Q}^{2}}l_{0}^{2}}{2}}}
\end{equation}
Excluding the intercalations, the Hamiltonian of the Coulomb interaction in the presence of the Landau quantization and Rashba spin-orbit coupling has the form:
\begin{widetext}
\begin{eqnarray}
&{{H}_{Coul}}=\frac{1}{2}\sum\limits_{{\vec{Q}}}{{}}W(\vec{Q})\{\sum\limits_{i,j}{{}}S({{R}_{i}};\vec{Q})S({{R}_{j}};-\vec{Q})[\hat{\rho }({{R}_{i}};\vec{Q})\hat{\rho }({{R}_{j}};-\vec{Q})-{{\delta }_{i,j}}\hat{\rho }({{R}_{i}};0)]+& \nonumber \\
&+\sum\limits_{{{M}_{v}},{{{{M}'}}_{v}}}{\sum\limits_{{{\varepsilon }_{m}},\varepsilon _{{{m}'}}^{{}}}{{}}}S({{M}_{v}},{{\varepsilon }_{m}};\vec{Q})S({{{{M}'}}_{v}},{{\varepsilon }_{{{m}'}}};-\vec{Q})[\hat{\rho }({{M}_{v}},{{\varepsilon }_{m}};\vec{Q})\hat{\rho }({{{{M}'}}_{v}},{{\varepsilon }_{{{m}'}}};-\vec{Q})-{{\delta }_{{{M}_{v}},{{{{M}'}}_{v}}}}{{\delta }_{m,{m}'}}\hat{\rho }({{M}_{v}},{{\varepsilon }_{m}};0)]+& \nonumber \\
&+\sum\limits_{{{M}_{v}},{{{{M}'}}_{v}}}{\sum\limits_{\varepsilon _{m}^{-},\varepsilon _{{{m}'}}^{-}}{{}}}S({{M}_{v}},\varepsilon _{m}^{-};\vec{Q})S({{{{M}'}}_{v}},\varepsilon _{{{m}'}}^{-};-\vec{Q})[\hat{\rho }({{M}_{v}},\varepsilon _{m}^{-};\vec{Q})\hat{\rho }({{{{M}'}}_{v}},\varepsilon _{{{m}'}}^{-};-\vec{Q})-{{\delta }_{{{M}_{v}},{{{{M}'}}_{v}}}}{{\delta }_{m,{m}'}}\hat{\rho }({{M}_{v}},\varepsilon _{m}^{-};0)]+ &\nonumber \\
&+\sum\limits_{i=1,2}{\sum\limits_{{{M}_{v}},{{\varepsilon }_{m}}}{{}}}S({{R}_{i}};\vec{Q})S({{M}_{v}},{{\varepsilon }_{m}};-\vec{Q})\hat{\rho }({{R}_{i}};\vec{Q})\hat{\rho }({{M}_{v}},{{\varepsilon }_{m}};-\vec{Q})+ &\nonumber \\
&+\sum\limits_{i=1,2}{\sum\limits_{{{M}_{v}},\varepsilon _{m}^{-}}{{}}}S({{R}_{i}};\vec{Q})S({{M}_{v}},\varepsilon _{m}^{-};-\vec{Q})\hat{\rho }({{R}_{i}};\vec{Q})\hat{\rho }({{M}_{v}},\varepsilon _{m}^{-};-\vec{Q})+ &\nonumber \\
&+\sum\limits_{i=1,2}{\sum\limits_{{{M}_{v}},{{\varepsilon }_{m}}}{{}}}S({{M}_{v}},{{\varepsilon }_{m}};\vec{Q})S({{R}_{i}};-\vec{Q})\hat{\rho }({{M}_{v}},{{\varepsilon }_{m}};\vec{Q})\hat{\rho }({{R}_{i}};-\vec{Q})+ &\nonumber \\
&+\sum\limits_{i=1,2}{\sum\limits_{{{M}_{v}},\varepsilon _{m}^{-}}{{}}}S({{M}_{v}},\varepsilon _{m}^{-};\vec{Q})S({{R}_{i}};-\vec{Q})\hat{\rho }({{M}_{v}},\varepsilon _{m}^{-};\vec{Q})\hat{\rho }({{R}_{i}};-\vec{Q})+ &\nonumber \\
&+\sum\limits_{{{M}_{v}},{{{{M}'}}_{v}}}{\sum\limits_{\varepsilon _{m}^{{}},\varepsilon _{{{m}'}}^{-}}{{}}}S({{M}_{v}},{{\varepsilon }_{m}};\vec{Q})S({{{{M}'}}_{v}},\varepsilon _{{{m}'}}^{-};-\vec{Q})\hat{\rho }({{M}_{v}},{{\varepsilon }_{m}};\vec{Q})\hat{\rho }({{{{M}'}}_{v}},\varepsilon _{{{m}'}}^{-};-\vec{Q})+ &\nonumber \\
&+\sum\limits_{{{M}_{v}},{{{{M}'}}_{v}}}{\sum\limits_{\varepsilon _{m}^{-},\varepsilon _{{{m}'}}^{{}}}{{}}}S({{M}_{v}},\varepsilon _{m}^{-};\vec{Q})S({{{{M}'}}_{v}},{{\varepsilon }_{{{m}'}}};-\vec{Q})\hat{\rho }({{M}_{v}},\varepsilon _{m}^{-};\vec{Q})\hat{\rho }({{{{M}'}}_{v}},{{\varepsilon }_{{{m}'}}};-\vec{Q})\}&
\end{eqnarray}
The Hamiltonian of the Coulomb interaction in the electron-hole representation looks as
\begin{eqnarray}
&{{H}_{Coul}}=\frac{1}{2}\sum\limits_{{\vec{Q}}}{{}}W(\vec{Q})\{\sum\limits_{i,j}{{}}S({{R}_{i}};\vec{Q})S({{R}_{j}};-\vec{Q})[{{{\hat{\rho }}}_{e}}({{R}_{i}};\vec{Q}){{{\hat{\rho }}}_{e}}({{R}_{j}};-\vec{Q})-{{\delta }_{i,j}}{{{\hat{\rho }}}_{e}}({{R}_{i}};0)]+ &\nonumber \\
&+\sum\limits_{{{M}_{h}},{{{{M}'}}_{h}}}{\sum\limits_{{{\varepsilon }_{m}},\varepsilon _{{{m}'}}^{{}}}{{}}}S({{M}_{h}},{{\varepsilon }_{m}};\vec{Q})S({{{{M}'}}_{h}},{{\varepsilon }_{{{m}'}}};-\vec{Q})[{{{\hat{\rho }}}_{h}}({{M}_{h}},{{\varepsilon }_{m}};\vec{Q}){{{\hat{\rho }}}_{h}}({{{{M}'}}_{h}},{{\varepsilon }_{{{m}'}}};-\vec{Q})-{{\delta }_{{{M}_{h}},{{{{M}'}}_{h}}}}{{\delta }_{m,{m}'}}{{{\hat{\rho }}}_{h}}({{M}_{h}},{{\varepsilon }_{m}};0)]+ &\nonumber \\
&+\sum\limits_{{{M}_{h}},{{{{M}'}}_{h}}}{\sum\limits_{\varepsilon _{m}^{-},\varepsilon _{{{m}'}}^{-}}{{}}}S({{M}_{h}},\varepsilon _{m}^{-};\vec{Q})S({{{{M}'}}_{h}},\varepsilon _{{{m}'}}^{-};-\vec{Q})[{{{\hat{\rho }}}_{h}}({{M}_{h}},\varepsilon _{m}^{-};\vec{Q}){{{\hat{\rho }}}_{h}}({{{{M}'}}_{h}},\varepsilon _{{{m}'}}^{-};-\vec{Q})-{{\delta }_{{{M}_{h}},{{{{M}'}}_{h}}}}{{\delta }_{m,{m}'}}{{{\hat{\rho }}}_{h}}({{M}_{h}},\varepsilon _{m}^{-};0)]+& \nonumber \\
&+\sum\limits_{{{M}_{h}},{{{{M}'}}_{h}}}{\sum\limits_{\varepsilon _{m}^{{}},\varepsilon _{{{m}'}}^{-}}{{}}}S({{M}_{h}},{{\varepsilon }_{m}};\vec{Q})S({{{{M}'}}_{h}},\varepsilon _{{{m}'}}^{-};-\vec{Q}){{{\hat{\rho }}}_{h}}({{M}_{h}},{{\varepsilon }_{m}};\vec{Q}){{{\hat{\rho }}}_{h}}({{{{M}'}}_{h}},\varepsilon _{{{m}'}}^{-};-\vec{Q})+& \nonumber \\
&+\sum\limits_{{{M}_{h}},{{{{M}'}}_{h}}}{\sum\limits_{\varepsilon _{m}^{-},\varepsilon _{{{m}'}}^{{}}}{{}}}S({{M}_{h}},\varepsilon _{m}^{-};\vec{Q})S({{{{M}'}}_{h}},{{\varepsilon }_{{{m}'}}};-\vec{Q}){{{\hat{\rho }}}_{h}}({{M}_{h}},\varepsilon _{m}^{-};\vec{Q}){{{\hat{\rho }}}_{h}}({{{{M}'}}_{h}},{{\varepsilon }_{{{m}'}}};-\vec{Q})- & \nonumber \\
&-\sum\limits_{i}{\sum\limits_{{{M}_{h}},{{\varepsilon }_{m}}}{{}}}S({{R}_{i}};\vec{Q})S({{M}_{h}},{{\varepsilon }_{m}};-\vec{Q}){{{\hat{\rho }}}_{e}}({{R}_{i}};\vec{Q}){{{\hat{\rho }}}_{h}}({{M}_{h}},{{\varepsilon }_{m}};-\vec{Q})- & \nonumber \\
&-\sum\limits_{i}{\sum\limits_{{{M}_{h}},\varepsilon _{m}^{-}}{{}}}S({{R}_{i}};\vec{Q})S({{M}_{h}},\varepsilon _{m}^{-};-\vec{Q}){{{\hat{\rho }}}_{e}}({{R}_{i}};\vec{Q}){{{\hat{\rho }}}_{h}}({{M}_{h}},\varepsilon _{m}^{-};-\vec{Q})- &\nonumber \\
&-\sum\limits_{i}{\sum\limits_{{{M}_{h}},{{\varepsilon }_{m}}}{{}}}S({{M}_{h}},{{\varepsilon }_{m}};\vec{Q})S({{R}_{i}};-\vec{Q}){{{\hat{\rho }}}_{h}}({{M}_{h}},{{\varepsilon }_{m}};\vec{Q}){{{\hat{\rho }}}_{e}}({{R}_{i}};-\vec{Q})- & \nonumber \\
&-\sum\limits_{i}{\sum\limits_{{{M}_{h}},\varepsilon _{m}^{-}}{{}}}S({{M}_{h}},\varepsilon _{m}^{-};\vec{Q})S({{R}_{i}};-\vec{Q}){{{\hat{\rho }}}_{h}}({{M}_{h}},\varepsilon _{m}^{-};\vec{Q}){{{\hat{\rho }}}_{e}}({{R}_{i}};-\vec{Q})\}&
\end{eqnarray}
In the concrete variant named as ${{F}_{1}}$, when the electrons are in the state ${{R}_{1}}$, whereas the holes are in the state $\varepsilon _{3}^{-}$ with a given value of ${{M}_{h}}$ the Hamiltonian (66) looks as
\begin{eqnarray}
& {{H}_{Coul}}({{R}_{1}};\varepsilon _{3}^{-})=\frac{1}{2}\sum\limits_{{\vec{Q}}}{{}}W(\vec{Q})\{{{(|{{a}_{0}}{{|}^{2}}+|{{b}_{1}}{{|}^{2}}{{A}_{1,1}}(\vec{Q}))}^{2}}[{{{\hat{\rho }}}_{e}}({{R}_{1}};\vec{Q}){{{\hat{\rho }}}_{e}}({{R}_{1}};-\vec{Q})-{{{\hat{\rho }}}_{e}}({{R}_{1}};0)]+ & \nonumber\\
& +{{(|d_{0}^{-}{{|}^{2}}+|c_{3}^{-}{{|}^{2}}{{A}_{3,3}}(\vec{Q}))}^{2}}[{{{\hat{\rho }}}_{h}}({{M}_{h}},\varepsilon _{3}^{-};\vec{Q}){{{\hat{\rho }}}_{h}}({{M}_{h}},\varepsilon _{3}^{-};-\vec{Q})-{{{\hat{\rho }}}_{h}}({{M}_{h}},\varepsilon _{3}^{-};0)]- & \nonumber\\
& -2(|{{a}_{0}}{{|}^{2}}+|{{b}_{1}}{{|}^{2}}{{A}_{1,1}}(\vec{Q}))(|d_{0}^{-}{{|}^{2}}+|c_{3}^{-}{{|}^{2}}{{A}_{3,3}}(\vec{Q})){{{\hat{\rho }}}_{e}}({{R}_{1}};\vec{Q}){{{\hat{\rho }}}_{h}}({{M}_{h}},\varepsilon _{3}^{-};-\vec{Q})\} &
\end{eqnarray}
\end{widetext}
In the absence of the RSOC we have ${{a}_{0}}=d_{0}^{-}=1$ and ${{b}_{1}}=c_{3}^{-}=0$. In the variant ${{F}_{1}}=({{R}_{1}},\varepsilon _{3}^{-})$ described by the Hamiltonian (67) the 2D magnetoexciton can be described by the wave function
\begin{equation}
\left| {{\Psi }_{ex}}({{F}_{1}},\vec{K}) \right\rangle =\frac{1}{\sqrt{N}}\sum\limits_{t}{{}}{{e}^{i{{K}_{y}}tl_{0}^{2}}}a_{{{R}_{1}},\frac{{{K}_{x}}}{2}+t}^{\dagger }b_{{{M}_{h}},\varepsilon _{3}^{-},\frac{{{K}_{x}}}{2}-t}^{\dagger }\left| 0 \right\rangle
\end{equation}
where $\left| 0 \right\rangle $ is the vacuum state determined by the equalities
\begin{equation}
{{a}_{\xi ,t}}\left| 0 \right\rangle ={{b}_{\eta ,t}}\left| 0 \right\rangle =0
\end{equation}
In the Ref.[22] there were considered also another seven combinations of the electron and hole states as follows:
\begin{eqnarray}
{{F}_{2}}=({{R}_{2}},\varepsilon _{3}^{-}),{{F}_{3}}=({{R}_{1}},\varepsilon _{0}), \nonumber \\
{{F}_{4}}=({{R}_{2}},\varepsilon _{0}^{{}}),{{F}_{5}}=({{R}_{1}},\varepsilon _{4}^{-} ), \nonumber \\
{{F}_{6}}=({{R}_{2}},\varepsilon _{4}^{-}),{{F}_{7}}=({{R}_{1}},\varepsilon _{1}), \nonumber \\
{{F}_{8}}=({{R}_{2}},\varepsilon _{1})
\end{eqnarray}
In all these cases the exciton creation energies were calculated using the formulas
\begin{eqnarray}
& {{E}_{ex}}({{F}_{n}},\vec{k})={{E}_{cv}}({{F}_{n}})-{{I}_{ex}}({{F}_{n}},\vec{k})& \nonumber  \\
& {{E}_{cv}}({{F}_{n}})-{{E}_{g}}={{E}_{e}}(\xi )+{{E}_{h}}(\eta ),& \nonumber \\
& {{F}_{n}}=(\xi ,\eta )&
\end{eqnarray}
Here ${{E}_{g}}$ is the semiconductor energy gap in the absence of the magnetic field. ${{I}_{ex}}({{F}_{n}},\vec{k})$ is the ionization potential of the magnetoexciton.

The Hamiltonian of the Coulomb electron-electron interaction in the case of e-h pairs with the electrons in the degenerate state $({{S}_{e}},{{R}_{1}})$ and the holes in the degenerate state $({{S}_{h}},{{M}_{h}},\varepsilon _{m}^{-})$ has the form
\begin{widetext}
\begin{eqnarray}
&{{H}_{Coul}}({{R}_{1}},\varepsilon _{m}^{-})= &\\
&=\frac{1}{2}\sum\limits_{{\vec{Q}}}{{}}\left\{ {{W}_{e-e}}({{R}_{1}};\vec{Q})\left[ {{{\hat{\rho }}}_{e}}({{S}_{e}},{{R}_{1}};\vec{Q}) \right. \right.{{{\hat{\rho }}}_{e}}({{S}_{e}},{{R}_{1}};-\vec{Q})-\left. {{{\hat{N}}}_{e}}({{S}_{e}},{{R}_{1}}) \right]+ &\nonumber  \\
&+{{W}_{h-h}}(\varepsilon _{m}^{-};\vec{Q})\left[ {{{\hat{\rho }}}_{h}}({{S}_{h}},{{M}_{h}},\varepsilon _{m}^{-};\vec{Q}){{{\hat{\rho }}}_{h}}({{S}_{h}},{{M}_{h}},\varepsilon _{m}^{-};-\vec{Q})-{{{\hat{N}}}_{h}}({{S}_{h}},{{M}_{h}},\varepsilon _{m}^{-}) \right]-& \nonumber \\
&-\left. 2{{W}_{e-h}}({{R}_{1}},\varepsilon _{m}^{-};\vec{Q}){{{\hat{\rho }}}_{e}}({{S}_{e}},{{R}_{1}};\vec{Q}){{{\hat{\rho }}}_{h}}({{S}_{h}},{{M}_{h}},\varepsilon _{m}^{-};-\vec{Q}) \right\} &\nonumber
\end{eqnarray}
\end{widetext}
Once again the electron and hole density operators are remembered
\begin{eqnarray}
&{{{\hat{\rho }}}_{e}}({{S}_{e}},{{R}_{1}};\vec{Q})=\sum\limits_{t}{{{e}^{i{{Q}_{y}}tl_{0}^{2}}}a_{{{S}_{e}},{{R}_{1}},t+\frac{{{Q}_{x}}}{2}}^{\dagger }a_{{{S}_{e}},{{R}_{1}},t-\frac{{{Q}_{x}}}{2}}^{{}}},& \nonumber\\
&{{{\hat{\rho }}}_{h}}({{S}_{h}},{{M}_{h}},\varepsilon _{m}^{-};\vec{Q})=\sum\limits_{t}{{{e}^{-i{{Q}_{y}}tl_{0}^{2}}}b_{{{S}_{h}},{{M}_{h}},\varepsilon _{m}^{-},t+\frac{{{Q}_{x}}}{2}}^{\dagger }b_{{{S}_{h}},{{M}_{h}},\varepsilon _{m}^{-},t-\frac{{{Q}_{x}}}{2}}^{{}}},& \nonumber\\
&{{{\hat{N}}}_{e}}({{S}_{e}},{{R}_{1}})={{{\hat{\rho }}}_{e}}({{S}_{e}},{{R}_{1}};0),& \\
&{{{\hat{N}}}_{h}}({{S}_{h}},{{M}_{h}},\varepsilon _{m}^{-})={{{\hat{\rho }}}_{h}}({{S}_{h}},{{M}_{h}},\varepsilon _{m}^{-};0)&\nonumber
\end{eqnarray}
The coefficients ${{W}_{i-j}}(\vec{Q})$ in (72) are
\begin{eqnarray}
&{{W}_{e-e}}({{R}_{1}};\vec{Q})=W(\vec{Q}){{\left( {{\left| {{a}_{0}} \right|}^{2}}{{A}_{0,0}}(\vec{Q})+{{\left| {{b}_{1}} \right|}^{2}}{{A}_{1,1}}(\vec{Q}) \right)}^{2}}, & \\
&{{W}_{h-h}}(\varepsilon _{m}^{-};\vec{Q})=W(\vec{Q}){{\left( {{\left| d_{m-3}^{-} \right|}^{2}}{{A}_{m-3,m-3}}(\vec{Q})+{{\left| c_{m}^{-} \right|}^{2}}{{A}_{m,m}}(\vec{Q}) \right)}^{2}}, &\nonumber \\
&m\ge 3,& \nonumber \\
&{{W}_{e-h}}({{R}_{1}},\varepsilon _{m}^{-};\vec{Q})=W(\vec{Q})\left( {{\left| {{a}_{0}} \right|}^{2}}{{A}_{0,0}}(\vec{Q})+{{\left| {{b}_{1}} \right|}^{2}}{{A}_{1,1}}(\vec{Q}) \right)\times & \nonumber \\
&\times \left( {{\left| d_{m-3}^{-} \right|}^{2}}{{A}_{m-3,m-3}}(\vec{Q})+{{\left| c_{m}^{-} \right|}^{2}}{{A}_{m,m}}(\vec{Q}) \right),& \nonumber \\
&m\ge 3 &\nonumber
\end{eqnarray}
The normalization conditions take place
\begin{eqnarray}
&{{\left| {{a}_{0}} \right|}^{2}}+{{\left| {{b}_{1}} \right|}^{2}}=1, & \nonumber \\
&{{\left| d_{m-3}^{-} \right|}^{2}}+{{\left| c_{m}^{-} \right|}^{2}}=1, & \nonumber \\
& m\ge 3&
\end{eqnarray}
In the actual case $m=3$ we obtained
\begin{eqnarray}
&{{W}_{e-e}}({{R}_{1}};\vec{Q})=W(\vec{Q}){{\left( 1-\frac{{{\left| {{b}_{1}} \right|}^{2}}}{2}{{Q}^{2}}l_{0}^{2} \right)}^{2}},&  \\
& {{W}_{h-h}}(\varepsilon _{3}^{-};\vec{Q})=W(\vec{Q})\times & \nonumber \\
&\times{{\left( 1-\frac{{{\left| c_{3}^{-} \right|}^{2}}}{2}\left( 3{{Q}^{2}}l_{0}^{2}-\frac{3}{4}{{Q}^{4}}l_{0}^{4}+\frac{1}{24}{{Q}^{6}}l_{0}^{6} \right) \right)}^{2}}, & \nonumber \\
& {{W}_{e-h}}({{R}_{1}},\varepsilon _{3}^{-};\vec{Q})=W(\vec{Q})\left( 1-\frac{{{\left| {{b}_{1}} \right|}^{2}}}{2}{{Q}^{2}}l_{0}^{2} \right)\times  & \nonumber\\
& \times \left( 1-\frac{{{\left| c_{3}^{-} \right|}^{2}}}{2}\left( 3{{Q}^{2}}l_{0}^{2}-\frac{3}{4}{{Q}^{4}}l_{0}^{4}+\frac{1}{24}{{Q}^{6}}l_{0}^{6} \right) \right), & \nonumber\\
& W(\vec{Q})={{e}^{-\frac{{{Q}^{2}}l_{0}^{2}}{2}}}V(\vec{Q}),\text{  }V(\vec{Q})=\frac{2\pi {{e}^{2}}}{{{\varepsilon }_{0}}S\left| {\vec{Q}} \right|}& \nonumber
\end{eqnarray}
The terms in (72) proportional to ${{\hat{N}}_{e}}$ and ${{\hat{N}}_{h}}$  have the coefficients which determine the partial ionization potentials for the electron ${{I}_{e}}({{R}_{1}})$ and for the hole ${{I}_{h}}(\varepsilon _{m}^{-})$ when they are bound in the ground state of the magnetoexciton with wave vector equal to zero. Their sum equals to the corresponding magnetoexciton ionization potential. For comparison the ionization potential in the absence of the RSOC is added. They are:
\begin{eqnarray}
&{{I}_{e}}({{R}_{1}})=\frac{1}{2}\sum\limits_{{\vec{Q}}}{{{W}_{e-e}}({{R}_{1}};\vec{Q})},& \nonumber \\
& {{I}_{h}}(\varepsilon _{m}^{-})=\frac{1}{2}\sum\limits_{{\vec{Q}}}{{{W}_{h-h}}(\varepsilon _{m}^{-};\vec{Q})},& \nonumber \\
& {{I}_{l}}({{R}_{1}},\varepsilon _{m}^{-})={{I}_{e}}({{R}_{1}})+{{I}_{h}}(\varepsilon _{m}^{-}), & \\
& I_{l}=\sum\limits_{\vec{Q}}W(\vec{Q})=\lim\limits_{\vec{P}\rightarrow\infty}E(\vec{P})= & \nonumber \\
& =\lim\limits_{\vec{P}\rightarrow\infty}2\sum\limits_{{\vec{Q}}}{W(\vec{Q})}{{\sin }^{2}}\left( \frac{{{\left[ \vec{P}\times \vec{Q} \right]}_{z}}l_{0}^{2}}{2} \right) & \nonumber
\end{eqnarray}
The terms of the Hamiltonian, which contain only the operators ${{\hat{N}}_{e}}$ and ${{\hat{N}}_{h}}$ describing the full number of electrons and holes including their chemical potentials are gathered in a form
\begin{eqnarray}
& {{H}_{mex,1}}=\left( {{E}_{e}}({{S}_{e}},{{R}_{1}})-{{I}_{e}}({{R}_{1}})-{{\mu }_{e}} \right){{{\hat{N}}}_{e}}({{S}_{e}},{{R}_{1}})+& \\
& +\left( {{E}_{h}}({{S}_{h}},{{M}_{h}},\varepsilon _{m}^{-})-{{I}_{h}}(\varepsilon _{m}^{-})-{{\mu }_{h}} \right){{{\hat{N}}}_{h}}({{S}_{h}},{{M}_{h}},\varepsilon _{m}^{-})& \nonumber
\end{eqnarray}
Now instead of the electron and hole density operators ${{\hat{\rho }}_{e}}(\vec{Q})$ and ${{\hat{\rho }}_{h}}(\vec{Q})$ we will introduce the density operators of the optical plasmon denoted as $\hat{\rho }(\vec{Q})$ and of the acoustical plasmon denoted ad $\hat{D}(\vec{Q})$ following the relations
\begin{eqnarray}
& \hat{\rho }(\vec{Q})={{{\hat{\rho }}}_{e}}(\vec{Q})-{{{\hat{\rho }}}_{h}}(\vec{Q}),& \nonumber \\
& \hat{D}(\vec{Q})={{{\hat{\rho }}}_{e}}(\vec{Q})+{{{\hat{\rho }}}_{h}}(\vec{Q}), & \nonumber  \\
& {{{\hat{\rho }}}_{e}}(\vec{Q})=\frac{\hat{\rho }(\vec{Q})+\hat{D}(\vec{Q})}{2}, & \nonumber  \\
& {{{\hat{\rho }}}_{h}}(\vec{Q})=\frac{\hat{D}(\vec{Q})-\hat{\rho }(\vec{Q})}{2} &
\end{eqnarray}
Here for simplicity many indexes, which label the electron, hole and plasmon density operators, are omitted. But they must be kept in mind and may be restored in concrete cases.
In the plasmon representation the Hamiltonian ${{H}_{mex,1}}$ (78) looks as:
\begin{eqnarray}
& {{H}_{mex,1}}=\left( {{E}_{mex}}({{S}_{e}},{{R}_{1}};{{S}_{h}},{{M}_{h}},\varepsilon _{m}^{-})-{{\mu }_{mex}} \right)\frac{\hat{D}(0)}{2}+ &\nonumber \\
& +\left( {{G}_{e-h}}({{S}_{e}},{{R}_{1}};{{S}_{h}},{{M}_{h}},\varepsilon _{m}^{-})-{{\mu }_{e}}+{{\mu }_{h}} \right)\frac{\hat{\rho }(0)}{2} &
\end{eqnarray}
Here the sums and differences of the Landau quantization level energies, of the partial ionization potentials and of the chemical potentials are present as follows
\begin{eqnarray}
& {{E}_{mex}}({{S}_{e}},{{R}_{1}};{{S}_{h}},{{M}_{h}},\varepsilon _{m}^{-})= & \nonumber \\
& ={{E}_{g}}({{S}_{e}},{{R}_{1}};{{S}_{h}},{{M}_{h}},\varepsilon _{m}^{-})-{{I}_{l}}({{R}_{1}};\varepsilon _{m}^{-}), & \nonumber \\
& {{E}_{g}}({{S}_{e}},{{R}_{1}};{{S}_{h}},{{M}_{h}},\varepsilon _{m}^{-})= & \nonumber \\
& ={{E}_{e}}({{S}_{e}},{{R}_{1}})+{{E}_{h}}({{S}_{h}},{{M}_{h}},\varepsilon _{m}^{-}), & \nonumber \\
& {{\mu }_{ex}}={{\mu }_{e}}+{{\mu }_{h}}, & \\
& {{G}_{e-h}}({{S}_{e}},{{R}_{1}};{{S}_{h}},{{M}_{h}},\varepsilon _{m}^{-})= & \nonumber \\
& ={{E}_{e}}({{S}_{e}},{{R}_{1}})-{{E}_{h}}({{S}_{h}},{{M}_{h}},\varepsilon _{m}^{-})-{{I}_{e}}({{R}_{1}})+{{I}_{h}}(\varepsilon _{m}^{-}) & \nonumber
\end{eqnarray}
The remaining part ${{H}_{mex,2}}$ of the Hamiltonian (72) after the excluding of the linear terms is quadratic in the plasmon density operators. It has the form:
\begin{eqnarray}
& {{H}_{mex,2}}=\frac{1}{2}\sum\limits_{{\vec{Q}}}{\left\{ {{W}_{0-0}}(\vec{Q}) \right.}\hat{\rho }(\vec{Q})\hat{\rho }(-\vec{Q})+ & \nonumber \\
& +{{W}_{a-a}}(\vec{Q})\hat{D}(\vec{Q})\hat{D}(-\vec{Q})+ & \\
& \left. +{{W}_{0-a}}(\vec{Q})\left( \hat{\rho }(\vec{Q})\hat{D}(-\vec{Q})+\hat{D}(\vec{Q})\hat{\rho }(-\vec{Q}) \right) \right\}& \nonumber
\end{eqnarray}
The new coefficients are expressed through the former ones by the formulas
\begin{eqnarray}
& {{W}_{0-0}}(\vec{Q})=\frac{1}{4}\left( {{W}_{e-e}}(\vec{Q})+{{W}_{h-h}}(\vec{Q})+2{{W}_{e-h}}(\vec{Q}) \right), & \nonumber\\
& {{W}_{a-a}}(\vec{Q})=\frac{1}{4}\left( {{W}_{e-e}}(\vec{Q})+{{W}_{h-h}}(\vec{Q})-2{{W}_{e-h}}(\vec{Q}) \right), & \nonumber\\
& {{W}_{0-a}}(\vec{Q})=\frac{1}{4}\left( {{W}_{e-e}}(\vec{Q})-{{W}_{h-h}}(\vec{Q}) \right)&
\end{eqnarray}
In the case of the e-h pairs of the type $({{R}_{1}};\varepsilon _{m}^{-})$ they take the forms
\begin{eqnarray}
& {{W}_{0-0}}({{R}_{1}};\varepsilon _{m}^{-};\vec{Q})=\frac{1}{4}({{W}_{e-e}}({{R}_{1}};\vec{Q})+ & \nonumber\\
& +{{W}_{h-h}}(\varepsilon _{m}^{-};\vec{Q})+2{{W}_{e-h}}({{R}_{1}};\varepsilon _{m}^{-};\vec{Q}))= & \nonumber\\
& =\frac{W(\vec{Q})}{4}({{\left| {{a}_{0}} \right|}^{2}}{{A}_{0,0}}(\vec{Q})+{{\left| {{b}_{1}} \right|}^{2}}{{A}_{1,1}}(\vec{Q})+& \nonumber \\
& +{{\left| d_{m-3}^{-} \right|}^{2}}{{A}_{m-3,m-3}}(\vec{Q})+{{\left| c_{m}^{-} \right|}^{2}}{{A}_{m,m}}(\vec{Q}){{)}^{2}},& \nonumber \\
& {{W}_{a-a}}({{R}_{1}};\varepsilon _{m}^{-};\vec{Q})=\frac{1}{4}({{W}_{e-e}}({{R}_{1}};\vec{Q})+ & \nonumber\\
& +{{W}_{h-h}}(\varepsilon _{m}^{-};\vec{Q})-2{{W}_{e-h}}({{R}_{1}};\varepsilon _{m}^{-};\vec{Q}))= &\\
& =\frac{W(\vec{Q})}{4}({{\left| {{a}_{0}} \right|}^{2}}{{A}_{0,0}}(\vec{Q})+{{\left| {{b}_{1}} \right|}^{2}}{{A}_{1,1}}(\vec{Q})-& \nonumber \\
& -{{\left| d_{m-3}^{-} \right|}^{2}}{{A}_{m-3,m-3}}(\vec{Q})-{{\left| c_{m}^{-} \right|}^{2}}{{A}_{m,m}}(\vec{Q}){{)}^{2}}, & \nonumber\\
& {{W}_{0-a}}({{R}_{1}};\varepsilon _{m}^{-};\vec{Q})=\frac{1}{4}\left( {{W}_{e-e}}({{R}_{1}};\vec{Q})-{{W}_{h-h}}(\varepsilon _{m}^{-};\vec{Q}) \right)= & \nonumber\\
& =\frac{W(\vec{Q})}{4}[{{\left( {{\left| {{a}_{0}} \right|}^{2}}{{A}_{0,0}}(\vec{Q})+{{\left| {{b}_{1}} \right|}^{2}}{{A}_{1,1}}(\vec{Q}) \right)}^{2}}- & \nonumber\\
& -{{\left( {{\left| d_{m-3}^{-} \right|}^{2}}{{A}_{m-3,m-3}}(\vec{Q})+{{\left| c_{m}^{-} \right|}^{2}}{{A}_{m,m}}(\vec{Q}) \right)}^{2}}] & \nonumber
\end{eqnarray}
In a special case $m=3$ we have
\begin{eqnarray}
& {{W}_{0-0}}({{R}_{1}};\varepsilon _{m}^{-};\vec{Q})=W(\vec{Q})(1-\frac{{{\left| {{b}_{1}} \right|}^{2}}}{4}{{Q}^{2}}l_{0}^{2}- & \nonumber\\
& -\frac{{{\left| c_{m}^{-} \right|}^{2}}}{4}\left( 3{{Q}^{2}}l_{0}^{2}-\frac{3}{4}{{Q}^{4}}l_{0}^{4}+\frac{1}{24}{{Q}^{6}}l_{0}^{6} \right){{)}^{2}}, & \nonumber\\
& {{W}_{a-a}}({{R}_{1}};\varepsilon _{m}^{-};\vec{Q})=W(\vec{Q})(-\frac{{{\left| {{b}_{1}} \right|}^{2}}}{4}{{Q}^{2}}l_{0}^{2}+ & \nonumber\\
& +\frac{{{\left| c_{m}^{-} \right|}^{2}}}{4}\left( 3{{Q}^{2}}l_{0}^{2}-\frac{3}{4}{{Q}^{4}}l_{0}^{4}+\frac{1}{24}{{Q}^{6}}l_{0}^{6} \right){{)}^{2}}, &\\
& {{W}_{0-a}}({{R}_{1}};\varepsilon _{m}^{-};\vec{Q})=W(\vec{Q})[(1-\frac{{{\left| {{b}_{1}} \right|}^{2}}}{4}{{Q}^{2}}l_{0}^{2}- & \nonumber\\
& -\frac{{{\left| c_{m}^{-} \right|}^{2}}}{4}\left( 3{{Q}^{2}}l_{0}^{2}-\frac{3}{4}{{Q}^{4}}l_{0}^{4}+\frac{1}{24}{{Q}^{6}}l_{0}^{6} \right))\times & \nonumber \\
& \times (-\frac{{{\left| {{b}_{1}} \right|}^{2}}}{4}{{Q}^{2}}l_{0}^{2}+\frac{{{\left| c_{m}^{-} \right|}^{2}}}{4}\left( 3{{Q}^{2}}l_{0}^{2}-\frac{3}{4}{{Q}^{4}}l_{0}^{4}+\frac{1}{24}{{Q}^{6}}l_{0}^{6} \right))] & \nonumber
\end{eqnarray}
Side by side with the magnetoexciton subsystem, the photon subsystem does exist. In our case it is composed by the photons with a given circular polarization, for example, $\vec{\sigma }_{{\vec{k}}}^{+}$. Their wave vectors $\vec{k}={{\vec{a}}_{3}}\frac{\pi }{{{L}_{c}}}+{{\vec{k}}_{||}}$ have the same quantized longitudinal component equal to $\frac{\pi }{{{L}_{c}}}$, where ${{L}_{c}}$ is the resonator length and arbitrary values of the in-plane 2D vectors ${{\vec{k}}_{||}}$. The photon energies are $\hbar {{\omega }_{{\vec{k}}}}=\frac{\hbar c}{{{n}_{0}}}\sqrt{\frac{{{\pi }^{2}}}{L_{c}^{2}}+\vec{k}_{||}^{2}}$, where ${{n}_{0}}$ is the refractive index of the microcavity. The full number of the photons captured into the resonator is determined by their chemical potential ${{\mu }_{ph}}$.
The zeroth order Hamiltonian of the photons in microcavity looks as
\begin{equation}
{{H}_{0,ph}}=\sum\limits_{{{{\vec{k}}}_{||}}}{\left( \hbar {{\omega }_{{\vec{k}}}}-{{\mu }_{ph}} \right)}c_{\vec{k},\sigma }^{\dagger }c_{\vec{k},\sigma }^{{}}
\end{equation}
where $c_{\vec{k},\sigma }^{\dagger },c_{\vec{k},\sigma }^{{}}$ are the creation and annihilation photon operators and $\sigma $ denotes a definite circular polarization. Only the case $\sigma =-$ will be considered. It must be supplemented by the Hamiltonian of the magnetoexciton-photon interaction deduced above in more general case. In the case of dipole-active band-to-band quantum transition with the combination of the e-h states $({{R}_{1}},\varepsilon _{3}^{-})$ we have
\begin{eqnarray}
& {{H}_{mex-ph}}=\sum\limits_{{{{\vec{k}}}_{\parallel }}}{[\varphi ({{{\vec{k}}}_{||}};{{R}_{1}};\varepsilon _{3}^{-})}\left( \vec{\sigma }_{{\vec{k}}}^{+}\bullet \vec{\sigma }_{{{M}_{h}}}^{*} \right){{c}_{\vec{k},\sigma }}\hat{\Psi }_{ex}^{\dagger }({{{\vec{k}}}_{||}})+ & \nonumber\\
& +{{\varphi }^{*}}({{{\vec{k}}}_{||}};{{R}_{1}};\varepsilon _{3}^{-})\left( \vec{\sigma }_{{\vec{k}}}^{-}\bullet \vec{\sigma }_{{{M}_{h}}}^{{}} \right)c_{\vec{k},\sigma }^{\dagger }\hat{\Psi }_{ex}^{{}}({{{\vec{k}}}_{||}})]&
\end{eqnarray}
The interaction coefficient equals to the expression
\begin{eqnarray}
&\varphi ({{{\vec{k}}}_{||}};{{R}_{1}};\varepsilon _{3}^{-})=\left( -\frac{e}{{{m}_{0}}{{l}_{0}}} \right)\sqrt{\frac{\hbar }{{{L}_{c}}{{\omega }_{{\vec{k}}}}}}{{P}_{cv}}(0)T({{{\vec{k}}}_{||}};{{R}_{1}};\varepsilon _{3}^{-}), & \nonumber\\
&T({{{\vec{k}}}_{||}};{{R}_{1}};\varepsilon _{3}^{-})=a_{0}^{*}d_{0}^{-*}\tilde{\phi }(0,0;{{{\vec{k}}}_{||}})-b_{1}^{*}c_{3}^{-*}\tilde{\phi }(1,3;{{{\vec{k}}}_{||}})&
\end{eqnarray}
The magnetoexciton creation and annihilation operators were written in a shortened form in (87) because there are too many indexes in its full description as follows
\begin{eqnarray}
& \hat{\Psi }_{ex}^{\dagger }(\vec{Q})=\hat{\Psi }_{ex}^{\dagger }(\vec{Q};{{S}_{e}},{{R}_{1}};{{S}_{h}},{{M}_{h}},\varepsilon _{3}^{-})= & \nonumber\\
& =\frac{1}{\sqrt{N}}\sum\limits_{t}{{{e}^{i{{Q}_{y}}tl_{0}^{2}}}}a_{{{S}_{e}},{{R}_{1}},t+\frac{{{Q}_{x}}}{2}}^{\dagger }b_{{{S}_{h}},{{M}_{h}},\varepsilon _{3}^{-},-t+\frac{{{Q}_{x}}}{2}}^{\dagger }&
\end{eqnarray}
The full Hamiltonian of the magnetoexciton-photon system for a more actual combination $({{R}_{1}},\varepsilon _{3}^{-})$ may be written
\begin{equation}
H={{H}_{mex,1}}+{{H}_{0,ph}}+{{H}_{mex-ph}}+{{H}_{mex,2}}
\end{equation}
Its remarkable peculiarity is the presence only of the two-particle integral plasmon and magnetoexciton operators, rather than of the single-particle electron and hole Fermi operators. It permits to simplify considerably the deduction of their equations of motion. By this reason the commutation relations between the full set of four two-particle integral operators $\hat{\rho }(\vec{Q}),\hat{D}(\vec{Q}),\hat{\Psi }_{ex}^{\dagger }(\vec{Q})$ and $\hat{\Psi }_{ex}^{{}}(\vec{Q})$ are needed. They are listed below
\begin{eqnarray}
&\left[ \hat{\rho }(\vec{Q}),\hat{\rho }(\vec{P}) \right]=\left[ \hat{D}(\vec{Q}),\hat{D}(\vec{P}) \right]= & \nonumber\\
& =2i\sin \left( Z(\vec{P},\vec{Q}) \right)\hat{\rho }\left( \vec{Q}+\vec{P} \right), & \nonumber\\
& \left[ \hat{\rho }(\vec{Q}),\hat{D}(\vec{P}) \right]=2i\sin \left( Z(\vec{P},\vec{Q}) \right)\hat{D}\left( \vec{P}+\vec{Q} \right), & \nonumber\\
& \left[ {{{\hat{\Psi }}}_{ex}}(\vec{P}),\hat{\Psi }_{ex}^{\dagger }(\vec{Q}) \right]={{\delta }_{kr}}(\vec{P},\vec{Q})-& \nonumber \\
& -\frac{1}{N}[i\sin \left( Z(\vec{Q},\vec{P}) \right)\hat{\rho }\left( \vec{Q}-\vec{P} \right)+ & \nonumber\\
& +\cos \left( Z(\vec{Q},\vec{P}) \right)\hat{D}\left( \vec{Q}-\vec{P} \right)],& \nonumber \\
& Z(\vec{P},\vec{Q})=-Z(\vec{Q},\vec{P})=\frac{{{[\vec{P}\times \vec{Q}]}_{z}}l_{0}^{2}}{2} & \nonumber\\
& ==-Z(\vec{P},-\vec{Q})=Z(-\vec{Q},\vec{P}),& \nonumber \\
& \left[ {{{\hat{\Psi }}}_{ex}}(\vec{P}),\hat{\Psi }_{ex}^{\dagger }(\vec{P}) \right]=1-\frac{1}{N}\hat{D}(0), &\\
& \left[ \hat{\rho }(\vec{Q}),\hat{\Psi }_{ex}^{\dagger }(\vec{P}) \right]=2i\sin \left( Z(\vec{P},\vec{Q}) \right)\hat{\Psi }_{ex}^{\dagger }(\vec{P}+\vec{Q}),& \nonumber \\
& \left[ \hat{\rho }(\vec{Q}),\hat{\Psi }_{ex}^{{}}(\vec{P}) \right]=-2i\sin \left( Z(\vec{P},\vec{Q}) \right)\hat{\Psi }_{ex}^{{}}(\vec{P}-\vec{Q}),& \nonumber \\
& \left[ \hat{D}(\vec{Q}),\hat{\Psi }_{ex}^{\dagger }(\vec{P}) \right]=2\cos \left( Z(\vec{P},\vec{Q}) \right)\hat{\Psi }_{ex}^{\dagger }(\vec{P}+\vec{Q}), & \nonumber\\
& \left[ \hat{D}(\vec{Q}),\hat{\Psi }_{ex}^{{}}(\vec{P}) \right]=-2\cos \left( Z(\vec{P},\vec{Q}) \right)\hat{\Psi }_{ex}^{{}}(\vec{P}-\vec{Q})& \nonumber
\end{eqnarray}
\section{The magnetoexcitons in the Bose-gas model description}
The Hamiltonian describing the 2D e-h pairs with electrons and holes situated on the given Landau quantization levels and interacting between themselves through the Coulomb forces was represented as a sum ${{H}_{mex,1}}+{{H}_{mex,2}}$. It is expressed through the plasmon density operators $\hat{\rho }(\vec{Q})$ and $\hat{D}(\vec{Q})$. It is useful to represent it in the model of weakly interacting Bose gas. To archive this end the wave functions describing the free single magnetoexcitons $\left| {{\psi }_{mex}}(\vec{P}) \right\rangle $ as well as the pairs of the free magnetoexcitons with wave vectors $\vec{P}$ and $\vec{R}$ $\left| {{\psi }_{mex}}(\vec{P}),{{\psi }_{mex}}(\vec{R}) \right\rangle $ were introduces
\begin{eqnarray}
& \left| {{\psi }_{ex}}(\vec{P}) \right\rangle =\hat{\Psi }_{ex}^{\dagger }(\vec{P})\left| 0 \right\rangle , & \nonumber\\
& \left\langle  {{\psi }_{ex}}(\vec{P}) \right|=\left\langle  0 \right|\hat{\Psi }_{ex}^{{}}(\vec{P}), &\\
& \left| {{\psi }_{ex}}(\vec{P}),{{\psi }_{ex}}(\vec{R}) \right\rangle =\hat{\Psi }_{ex}^{\dagger }(\vec{P})\hat{\Psi }_{ex}^{\dagger }(\vec{R})\left| 0 \right\rangle , & \nonumber\\
& \left\langle  {{\psi }_{ex}}(\vec{P}),{{\psi }_{ex}}(\vec{R}) \right|=\left\langle  0 \right|\hat{\Psi }_{ex}^{{}}(\vec{R})\hat{\Psi }_{ex}^{{}}(\vec{P})& \nonumber
\end{eqnarray}
where $\left| 0 \right\rangle $ is the vacuum state of the semiconductor. They were used to calculate the matrix elements
\begin{eqnarray}
& {{E}_{mex}}(\vec{P})=\left\langle  {{\psi }_{ex}}(\vec{P}) \right|{{H}_{mex,1}}+{{H}_{mex,2}}\left| {{\psi }_{ex}}(\vec{P}) \right\rangle ,& \nonumber \\
& W({{{\vec{P}}}_{1}},{{{\vec{R}}}_{1}};{{{\vec{P}}}_{2}},{{{\vec{R}}}_{2}})= &\\
& =\left\langle  {{\psi }_{ex}}({{{\vec{P}}}_{1}}){{\psi }_{ex}}({{{\vec{R}}}_{1}}) \right|{{H}_{mex,1}}+{{H}_{mex,2}}\left| {{\psi }_{ex}}({{{\vec{P}}}_{2}}){{\psi }_{ex}}({{{\vec{R}}}_{2}}) \right\rangle & \nonumber
\end{eqnarray}
With these matrix elements and with the magnetoexciton creation and annihilation operators the new Hamiltonian in the model of the weakly interacting Bose gas can be constructed. It looks as
\begin{eqnarray}
& H={{H}_{0}}+{{H}_{int}}, & \nonumber \\
& {{H}_{0}}=\sum\limits_{{\vec{P}}}{{{E}_{mex}}(\vec{P})\hat{\Psi }_{ex}^{\dagger }}(\vec{P})\hat{\Psi }_{ex}^{{}}(\vec{P}), & \nonumber \\
& {{H}_{int}}=\sum\limits_{{{{\vec{P}}}_{1}},{{{\vec{R}}}_{1}},{{{\vec{P}}}_{2}},{{{\vec{R}}}_{2}}}{W({{{\vec{P}}}_{1}},{{{\vec{R}}}_{1}};{{{\vec{P}}}_{2}},{{{\vec{R}}}_{2}})}\times  &\\
& \times \hat{\Psi }_{ex}^{\dagger }({{{\vec{P}}}_{1}})\hat{\Psi }_{ex}^{\dagger }({{{\vec{R}}}_{1}})\hat{\Psi }_{ex}^{{}}({{{\vec{R}}}_{2}})\hat{\Psi }_{ex}^{{}}({{{\vec{P}}}_{2}}), & \nonumber\\
& {{{\vec{P}}}_{1}}+{{{\vec{R}}}_{1}}={{{\vec{P}}}_{2}}+{{{\vec{R}}}_{2}}& \nonumber
\end{eqnarray}
One can remember that the magnetoexciton creation and annihilation operators in their turn are constructed from the electron and hole creation and annihilation Fermi-type operators $a_{p}^{\dagger },{{a}_{p}},b_{p}^{\dagger },{{b}_{p}}$ as follows
\begin{equation}
\hat{\Psi }_{ex}^{\dagger }(\vec{P})=\frac{1}{\sqrt{N}}\sum\limits_{t}{{{e}^{i{{P}_{y}}tl_{0}^{2}}}}a_{t+\frac{{{P}_{x}}}{2}}^{\dagger }b_{-t+\frac{{{P}_{x}}}{2}}^{\dagger }
\end{equation}
and their composition in all calculations is taken into account. Some of them are demonstrated below using the commutation relations (91)
\begin{eqnarray}
& {{a}_{p}}\left| 0 \right\rangle ={{b}_{p}}\left| 0 \right\rangle =0, & \nonumber \\
& \hat{\rho }(\vec{Q})\left| 0 \right\rangle =\hat{D}(\vec{Q})\left| 0 \right\rangle ={{{\hat{\Psi }}}_{ex}}(\vec{Q})\left| 0 \right\rangle =0, & \nonumber \\
& \hat{D}(0)\hat{\Psi }_{ex}^{\dagger }(\vec{P})\left| 0 \right\rangle =2\hat{\Psi }_{ex}^{\dagger }(\vec{P})\left| 0 \right\rangle ,& \nonumber \\
& \hat{\rho }(0)\hat{\Psi }_{ex}^{\dagger }(\vec{P})\left| 0 \right\rangle =0, &\\
& \hat{\rho }(\vec{Q})\hat{\rho }(-\vec{Q})\hat{\Psi }_{ex}^{\dagger }(\vec{P})\left| 0 \right\rangle =4{{\sin }^{2}}(Z(\vec{P},\vec{Q}))\hat{\Psi }_{ex}^{\dagger }(\vec{P})\left| 0 \right\rangle , & \nonumber\\
& \hat{D}(\vec{Q})\hat{D}(-\vec{Q})\hat{\Psi }_{ex}^{\dagger }(\vec{P})\left| 0 \right\rangle =4{{\cos }^{2}}(Z(\vec{P},\vec{Q}))\hat{\Psi }_{ex}^{\dagger }(\vec{P})\left| 0 \right\rangle , & \nonumber \\
& (\hat{\rho }(\vec{Q})\hat{D}(-\vec{Q})+\hat{D}(\vec{Q})\hat{\rho }(-\vec{Q}))\hat{\Psi }_{ex}^{\dagger }(\vec{P})\left| 0 \right\rangle =0 & \nonumber
\end{eqnarray}
In the present model the main role is played by the magnetoexciton creation and annihilation operators, rather than by the plasmon density operators.

The magnetoexciton creation energy ${{E}_{mex}}(\vec{P})$ from the Hamiltonian ${{H}_{0}}$ consistes from three parts
\begin{eqnarray}
& {{E}_{mex}}(\vec{P})={{E}_{mex}}(\vec{P};{{S}_{e}},{{R}_{1}};{{S}_{h}},{{M}_{h}},\varepsilon _{m}^{-})= & \nonumber\\
& ={{E}_{g}}({{S}_{e}},{{R}_{1}};{{S}_{h}},{{M}_{h}},\varepsilon _{m}^{-})-{{I}_{l}}({{R}_{1}};\varepsilon _{m}^{-})+E(\vec{P};{{R}_{1}};\varepsilon _{m}^{-}), & \nonumber \\
& E(\vec{P};{{R}_{1}};\varepsilon _{m}^{-})=2\sum\limits_{{\vec{Q}}}{{{W}_{0-0}}(\vec{Q},{{R}_{1}};\varepsilon _{m}^{-}){{\sin }^{2}}(Z(\vec{P},\vec{Q}))}+ & \nonumber \\
& +2\sum\limits_{{\vec{Q}}}{{{W}_{a-a}}(\vec{Q},{{R}_{1}};\varepsilon _{m}^{-}){{\cos }^{2}}(Z(\vec{P},\vec{Q}))} &
\end{eqnarray}
The first component ${{E}_{g}}({{S}_{e}},{{R}_{1}};{{S}_{h}},{{M}_{h}},\varepsilon _{m}^{-})$ plays the role of the band gap, whereas the difference ${{I}_{l}}({{R}_{1}};\varepsilon _{m}^{-})-E(\vec{P};{{R}_{1}};\varepsilon _{m}^{-})$ determines the resulting ionization potential of the moving magnetoexciton with wave vector $\vec{P}$. In the limiting case $\vec{P}\to \infty $, when $\lim\limits_{\vec{P}\to \infty }E(\vec{P};{{R}_{1}};\varepsilon _{m}^{-})={{I}_{l}}({{R}_{1}};\varepsilon _{m}^{-})$, the resulting ionization potential vanishes and the e-h pair becomes unbound. Nevertheless the presence of the positive term $E(\vec{P};{{R}_{1}};\varepsilon _{m}^{-})$ in the formula (97) plays the role of the kinetic energy of the magnetoexciton at least in the region of the small values of the wave vector $\vec{P}$, where this term can be represented in a quadratic form $\frac{{{\hbar }^{2}}{{P}^{2}}}{2M(B)}$ with the effective mass $M(B)$ depending on the magnetic field strength B. The zeroth-order Hamiltonian ${{H}_{0}}$ (94) together with the similar Hamiltonian for the cavity photons and with the Hamiltonian describing the magnetoexciton-photon interaction give rise to the quadratic Hamiltonian ${{H}_{2}}$ forming the base of the polariton conception. It looks as
\begin{eqnarray}
& {{H}_{2}}=\sum\limits_{{{{\vec{k}}}_{||}}}{{{E}_{mex}}({{{\vec{k}}}_{||}})\hat{\Psi }_{ex}^{\dagger }({{{\vec{k}}}_{||}})\hat{\Psi }_{ex}^{{}}({{{\vec{k}}}_{||}})}+ & \nonumber \\
& +\sum\limits_{{{{\vec{k}}}_{||}}}{\hbar {{\omega }_{{\vec{k}}}}c_{\vec{k},-}^{\dagger }{{c}_{\vec{k},-}}+} &\\
& +\sum\limits_{{{{\vec{k}}}_{||}}}{{}}[\varphi ({{{\vec{k}}}_{||}})(\vec{\sigma }_{{\vec{k}}}^{+}\bullet \vec{\sigma }_{{{M}_{h}}}^{*}){{c}_{\vec{k},-}}\hat{\Psi }_{ex}^{\dagger }({{{\vec{k}}}_{||}})+ & \nonumber\\
& +{{\varphi }^{*}}({{{\vec{k}}}_{||}})(\vec{\sigma }_{{\vec{k}}}^{-}\bullet \vec{\sigma }_{{{M}_{h}}}^{{}})c_{\vec{k},-}^{\dagger }\hat{\Psi }_{ex}^{{}}({{{\vec{k}}}_{||}})] & \nonumber
\end{eqnarray}
In this expression the chemical potentials of the magnetoexcitons and of the photons are emitted up till the single-particle polariton formation is investigated. They will be restored whem the collective properties of the polaritons will be discussed. The diagonalization of the quadratic form (98) is achieved introducing the polariton creation and annihilation operators $\hat{L}_{{{{\vec{k}}}_{||}}}^{\dagger },\hat{L}_{{{{\vec{k}}}_{||}}}^{{}}$ in a form of a linear superposition
\begin{equation}
 \hat{L}_{{{{\vec{k}}}_{||}}}^{{}}=x({{\vec{k}}_{||}}){{\Psi }_{ex}}({{\vec{k}}_{||}})+y({{\vec{k}}_{||}}){{c}_{\vec{k},-}}
\end{equation}
It is a simplified form without the antiresonance terms because they are not introduced in the starting Hamiltonian ${{H}_{2}}$ (98). The values $x({{\vec{k}}_{||}})$ and $y({{\vec{k}}_{||}})$ are known as Hopfield coefficients. In the case when the scalar product of two circular polarized vectors equals 1, the energy spectrum of two polariton branches looks as
\begin{eqnarray}
& \hbar \omega ({{{\vec{k}}}_{||}})=\frac{{{E}_{mex}}({{{\vec{k}}}_{||}})+\hbar {{\omega }_{{\vec{k}}}}}{2}\pm  & \nonumber\\
& \pm \frac{1}{2}\sqrt{{{({{E}_{mex}}({{{\vec{k}}}_{||}})-\hbar {{\omega }_{{\vec{k}}}})}^{2}}+4|\varphi ({{{\vec{k}}}_{||}}){{|}^{2}}} &
\end{eqnarray}
The Rabi frequency for the e-h pair in the states $({{R}_{1}},\varepsilon _{3}^{-})$ equals to
\begin{eqnarray}
& |{{\omega }_{R}}|=\left| \frac{\varphi (0)}{\hbar } \right|= & \\
& =\frac{e}{{{m}_{0}}{{l}_{0}}}\sqrt{\frac{1}{{{L}_{c}}\hbar {{\omega }_{k}}}}\left| {{P}_{cv}}(0){{a}_{0}}{{d}_{0}} \right|\left| \vec{\sigma }_{{\vec{k}}}^{-}\bullet {{{\vec{\sigma }}}_{{{M}_{h}}}} \right| & \nonumber
\end{eqnarray}
In the absence of the RSOI the coefficients ${{a}_{0}}={{d}_{0}}=1$ and the expression (101) coincides with the formula (12) of the Ref[21].

The Hopfield coefficients obey to the normalization condition and equal to
\begin{eqnarray}
& {{\left| x({{{\vec{k}}}_{||}}) \right|}^{2}}=\frac{{{\left( \hbar {{\omega }_{p}}({{{\vec{k}}}_{||}})-\hbar {{\omega }_{{\vec{k}}}} \right)}^{2}}}{{{\left( \hbar {{\omega }_{p}}({{{\vec{k}}}_{||}})-\hbar {{\omega }_{{\vec{k}}}} \right)}^{2}}+|\varphi ({{{\vec{k}}}_{||}})(\vec{\sigma }_{{\vec{k}}}^{-}\bullet \vec{\sigma }_{{{M}_{h}}}^{{}}){{|}^{2}}}, & \nonumber\\
& {{\left| y({{{\vec{k}}}_{||}}) \right|}^{2}}=\frac{|\varphi ({{{\vec{k}}}_{||}})(\vec{\sigma }_{{\vec{k}}}^{-}\bullet \vec{\sigma }_{{{M}_{h}}}^{{}}){{|}^{2}}}{{{\left( \hbar {{\omega }_{p}}({{{\vec{k}}}_{||}})-\hbar {{\omega }_{{\vec{k}}}} \right)}^{2}}+|\varphi ({{{\vec{k}}}_{||}})(\vec{\sigma }_{{\vec{k}}}^{-}\bullet \vec{\sigma }_{{{M}_{h}}}^{{}}){{|}^{2}}}, & \nonumber\\
& {{\left| x({{{\vec{k}}}_{||}}) \right|}^{2}}+{{\left| y({{{\vec{k}}}_{||}}) \right|}^{2}}=1, &\\
& \varphi ({{{\vec{k}}}_{||}})(\vec{\sigma }_{{\vec{k}}}^{+}\bullet \vec{\sigma }_{{{M}_{h}}}^{*})=|\varphi ({{{\vec{k}}}_{||}})(\vec{\sigma }_{{\vec{k}}}^{-}\bullet \vec{\sigma }_{{{M}_{h}}}^{{}})|{{e}^{i\gamma ({{{\vec{k}}}_{||}})}}, & \nonumber\\
& |(\vec{\sigma }_{{\vec{k}}}^{-}\bullet \vec{\sigma }_{{{M}_{h}}}^{{}})|=\left| (\vec{\sigma }_{{\vec{k}}}^{+}\bullet \vec{\sigma }_{{{M}_{h}}}^{*}) \right|, & \nonumber \\
& x({{{\vec{k}}}_{||}})=\left| x({{{\vec{k}}}_{||}}) \right|{{e}^{i\alpha ({{{\vec{k}}}_{||}})}},\text{ }y({{{\vec{k}}}_{||}})=\left| y({{{\vec{k}}}_{||}}) \right|{{e}^{i\beta ({{{\vec{k}}}_{||}})}}, & \nonumber \\
& \alpha ({{{\vec{k}}}_{||}})-\beta ({{{\vec{k}}}_{||}})+\gamma ({{{\vec{k}}}_{||}})=0 & \nonumber
\end{eqnarray}
The last equality results from the fact that the polariton energy spectrum $\hbar {{\omega }_{p}}({{\vec{k}}_{||}})$ as well as the magnetoexciton and cavity photon bare energies are real entities. This relation will be used below in the point ${{\vec{k}}_{||}}=0$ where these phases will be simply denoted as $\alpha ,\beta \text{ and }\gamma $.

Now the breaking of the gauge symmetry of the 2D magnetoexciton-photon system leading to the BEC of the magnetopolaritons on the lower polariton branch in the point ${{\vec{k}}_{||}}=0$ will be discussed.
\section{Breaking of the gauge symmetry and the mixed photon-magnetoexciton-acoustical plasmon states}
A method to introduce the coherent macroscopic polariton states in a system of 2D e-h pairs and photons captured in the microcavity was proposed in Refs[30, 31]. It was supposed that the e-h pairs were excited on the quantum well embedded into the microcavity and interact with the photons captured in the resonator giving rise to the 2D Wannier-Mott excitons and polariton formation. As was shown in [30] the proposed method is equivalent to the u-v Bogoliubov transformation for the electron and hole Fermi operators and to Bogoliubov displacement transformation for the photon Bose operators. Now this method will be applied to the case of 2D magnetoexcitons and photons in microcavity with the aim to investigate the BEC of magnetopolaritons in the state with ${{\vec{k}}_{||}}=0$ on the lower polariton branch. The unitary transformation proposed in Ref[30] looks as
\begin{equation}
D(\sqrt{{{N}_{p}}})=\exp \left( \sqrt{{{N}_{p}}}\left( L_{0}^{\dagger }-{{L}_{0}} \right) \right)
\end{equation}
where ${{N}_{p}}$ is a macroscopic number of the condensed polaritons in the point ${{\vec{k}}_{||}}=0$ of the lower polariton branch. The cavity photon with ${{\vec{k}}_{||}}=0$ has a quantized longitudinal projection of its wave vector $\vec{k}$ equal to $\pi /{{L}_{c}}$. Only the photons with a given circular polarization are considered. In this case we have
\begin{eqnarray}
& \hat{L}_{0}^{{}}=x(0){{\Psi }_{ex}}(0)+y(0){{c}_{\frac{\pi }{{{L}_{c}}},-}}, & \nonumber\\
& x(0)=\left| x(0) \right|{{e}^{i\alpha }}, & \nonumber\\
& y(0)=\left| y(0) \right|{{e}^{i\beta }} &
\end{eqnarray}
and the starting unitary transformation can be factorized in two independent unitary transformations acting separately in two subsystems of magnetoexcitons and of the photons as follows
\begin{eqnarray}
& D(\sqrt{{{N}_{p}}})= & \nonumber\\
& ={{D}_{ex}}(\sqrt{{{N}_{p}}}\left| x(0) \right|){{D}_{ph}}(\sqrt{{{N}_{p}}}\left| y(0) \right|), & \nonumber\\
& {{D}_{ex}}(\sqrt{{{N}_{p}}}\left| x(0) \right|)= &\\
& =\exp \left[ \sqrt{{{N}_{p}}}\left| x(0) \right|\left( {{e}^{-i\alpha }}\hat{\Psi }_{ex}^{\dagger }(0)-{{e}^{i\alpha }}\hat{\Psi }_{ex}^{{}}(0) \right) \right], & \nonumber\\
& {{D}_{ph}}(\sqrt{{{N}_{p}}}\left| y(0) \right|)= & \nonumber\\
& =\exp \left[ \sqrt{{{N}_{p}}}\left| y(0) \right|\left( {{e}^{-i\beta }}c_{\frac{\pi }{{{L}_{c}}},-}^{\dagger }-{{e}^{i\beta }}c_{\frac{\pi }{{{L}_{c}}},-}^{{}} \right) \right] & \nonumber
\end{eqnarray}
Taking into account the expressions for the magnetoexciton operators
\begin{eqnarray}
& \hat{\Psi }_{ex}^{\dagger }(0)=\frac{1}{\sqrt{N}}\sum\limits_{t}a_{t}^{\dagger }b_{-t}^{\dagger }, & \nonumber\\
& \hat{\Psi }_{ex}(0)=\frac{1}{\sqrt{N}}\sum\limits_{t}b_{-t}a_{t}&
\end{eqnarray}
one can transcribe the operator ${{D}_{ex}}(\sqrt{{{N}_{p}}}\left| x(0) \right|)$ in a form ${{D}_{ex}}(\sqrt{{N}_{p}}\left| x(0) \right|)={{e}^{z}}=\prod\limits_{t}{e}^{{z}_{t}}$, where
\begin{eqnarray}
& z=\sum\limits_{t}{{{z}_{t}}}= &\\
& =\sqrt{{{N}_{p}}}\left| x(0) \right|\left( {{e}^{-i\alpha }}\hat{\Psi }_{ex}^{\dagger }(0)-{{e}^{i\alpha }}\hat{\Psi }_{ex}^{{}}(0) \right), & \nonumber\\
& {{z}_{t}}={{\nu }_{p}}\left| x(0) \right|\left( {{e}^{-i\alpha }}a_{t}^{\dagger }b_{-t}^{\dagger }-{{e}^{i\alpha }}{{b}_{-t}}{{a}_{t}} \right) & \nonumber
\end{eqnarray}
The unitary transformations of the Fermi operator are
\begin{eqnarray}
& {{D}_{ex}}(\sqrt{{{N}_{p}}}\left| x(0) \right|){{a}_{t}}D_{ex}^{-1}(\sqrt{{{N}_{p}}}\left| x(0) \right|)= & \nonumber\\
& ={{e}^{{{z}_{t}}}}{{a}_{t}}{{e}^{-{{z}_{t}}}}={{\alpha }_{t}}= & \nonumber\\
& ={{a}_{t}}\cos ({{v}_{p}}\left| x(0) \right|)-b_{-t}^{\dagger }{{e}^{-i\alpha }}\sin ({{v}_{p}}\left| x(0) \right|), & \nonumber\\
& {{D}_{ex}}(\sqrt{{{N}_{p}}}\left| x(0) \right|){{b}_{-t}}D_{ex}^{-1}(\sqrt{{{N}_{p}}}\left| x(0) \right|)= & \nonumber\\
& ={{e}^{{{z}_{t}}}}{{b}_{-t}}{{e}^{-{{z}_{t}}}}={{\beta }_{-t}}= &\\
& ={{b}_{-t}}\cos ({{v}_{p}}\left| x(0) \right|)+a_{t}^{\dagger }{{e}^{-i\alpha }}\sin ({{v}_{p}}\left| x(0) \right|)& \nonumber
\end{eqnarray}
Here the filling factor of the Bose-Einstein condensate was introduced
\begin{equation}
\frac{{{N}_{p}}}{N}=v_{p}^{2}
\end{equation}
Side by side with the unitary transformations (108) for the single-particle Fermi operators, one can also obtain the transformations for the two-particle integral operators. They were obtained using the commutation relations (91) and look as follows
\begin{eqnarray}
& {{e}^{{\hat{z}}}}\frac{\hat{D}(\vec{Q})}{\sqrt{N}}{{e}^{-\hat{z}}}=\frac{\hat{D}(\vec{Q})}{\sqrt{N}}\cos (2{{v}_{p}}\left| x(0) \right|)- & \nonumber\\
& -\hat{\theta }(\vec{Q})\sin (2{{v}_{p}}\left| x(0) \right|), & \nonumber\\
& {{e}^{{\hat{z}}}}\hat{\theta }(\vec{Q}){{e}^{-\hat{z}}}=\hat{\theta }(\vec{Q})\cos (2{{v}_{p}}\left| x(0) \right|)+ & \nonumber\\
& +\frac{\hat{D}(\vec{Q})}{\sqrt{N}}\sin (2{{v}_{p}}\left| x(0) \right|), & \nonumber\\
& \hat{\theta }(\vec{Q})={{e}^{-i\alpha }}\hat{\Psi }_{ex}^{\dagger }(\vec{Q})+{{e}^{i\alpha }}\hat{\Psi }_{ex}^{{}}(-\vec{Q}), &\\
& {{e}^{{\hat{z}}}}\hat{\rho }(\vec{Q}){{e}^{-\hat{z}}}=\hat{\rho }(\vec{Q}), & \nonumber\\
& {{e}^{z}}{{e}^{-i\alpha }}\Psi _{ex}^{\dagger }(\vec{Q}){{e}^{-z}}={{e}^{-i\alpha }}\Psi _{ex}^{\dagger }(\vec{Q})+ & \nonumber\\
& +\frac{1}{2}\sin \left( 2{{v}_{p}}|x(0)| \right)\frac{D(\vec{Q})}{\sqrt{N}}+ & \nonumber\\
& +\frac{1}{2}[\cos \left( 2{{v}_{p}}|x(0)| \right)-1]\theta (\vec{Q}), & \nonumber\\
& {{e}^{z}}{{e}^{i\alpha }}\Psi _{ex}^{{}}(-\vec{Q}){{e}^{-z}}={{e}^{i\alpha }}\Psi _{ex}^{{}}(-\vec{Q})+ & \nonumber\\
& +\frac{1}{2}\sin \left( 2{{v}_{p}}|x(0)| \right)\frac{D(\vec{Q})}{\sqrt{N}}+ & \nonumber\\
& +\frac{1}{2}[\cos \left( 2{{v}_{p}}|x(0)| \right)-1]\theta (\vec{Q}) & \nonumber
\end{eqnarray}
As one can see the superposition of the magnetoexciton creation and annihilation operators in the form $\theta (\vec{Q})$ forms a coherent mixed state with the acoustical plasmon density opeartor $\frac{\hat{D}(\vec{Q})}{\sqrt{N}}$. Such mixed magnetoexciton-plasmon states were discussed in the Refs[32-34].

The full Hamiltonian of the magnetoexciton-photon system consists from four parts as follows
\begin{equation}
\hat{H}={{\hat{H}}_{mex,1}}+{{\hat{H}}_{0,ph}}+{{\hat{H}}_{mex,2}}+{{\hat{H}}_{mex-ph}}
\end{equation}
It will be subjected to the unitary gauge transformation (105), which means to calculate the following unitary transformations:
${{D}_{ex}}(\sqrt{{{N}_{p}}}|x(0)|)({{\hat{H}}_{mex,1}}+{{\hat{H}}_{mex,2}})D_{ex}^{-1}(\sqrt{{{N}_{p}}}|x(0)|)$, ${{D}_{ph}}(\sqrt{{{N}_{p}}}|y(0)|){{\hat{H}}_{0,ph}}D_{ph}^{-1}(\sqrt{{{N}_{p}}}|y(0)|)$, ${{D}_{ex}}(\sqrt{{{N}_{p}}}|x(0)|){{D}_{ph}}(\sqrt{{{N}_{p}}}|y(0)|){{\hat{H}}_{mex-ph}}\times$ $\times D_{ph}^{-1}(\sqrt{{{N}_{p}}}|y(0)|)D_{ex}^{-1}(\sqrt{{{N}_{p}}}|x(0)|)$.
The first of them is
\begin{eqnarray}
& {{{\hat{\tilde{H}}}}_{mex,1}}= & \nonumber \\
& ={{D}_{ex}}(\sqrt{{{N}_{p}}}|x(0)|){{{\hat{H}}}_{mex,1}}D_{ex}^{-1}(\sqrt{{{N}_{p}}}|x(0)|)= & \nonumber \\
& =\left( {{E}_{mex}}({{S}_{e}},{{R}_{1}};{{S}_{h}},{{M}_{h}},\varepsilon _{m}^{-})-{{\mu }_{ex}} \right)\times  & \nonumber \\
& \times \cos (2{{v}_{p}}|x(0)|)\frac{\hat{D}(0)}{2}+ & \\
& +({{G}_{e-h}}({{S}_{e}},{{R}_{1}};{{S}_{h}},{{M}_{h}},\varepsilon _{m}^{-})-{{\mu }_{e}}+{{\mu }_{h}})\frac{\hat{\rho }(0)}{2}- & \nonumber \\
& -\frac{\sqrt{N}}{2}\sin (2{{v}_{p}}|x(0)|)\times  & \nonumber \\
& \times \left( {{E}_{mex}}({{S}_{e}},{{R}_{1}};{{S}_{h}},{{M}_{h}},\varepsilon _{m}^{-})-{{\mu }_{ex}} \right)\hat{\theta }(0), & \nonumber \\
& {{\mu }_{ex}}={{\mu }_{e}}+{{\mu }_{h}} & \nonumber
\end{eqnarray}
The second one looks as
\begin{eqnarray}
& {{{\hat{\tilde{H}}}}_{mex,2}}= & \nonumber \\
& ={{D}_{ex}}(\sqrt{{{N}_{p}}}|x(0)|){{{\hat{H}}}_{mex,2}}D_{ex}^{-1}(\sqrt{{{N}_{p}}}|x(0)|)= & \nonumber \\
& =\frac{1}{2}\sum\limits_{{\vec{Q}}}{\left\{ {{W}_{0-0}}(\vec{Q})\hat{\rho }(\vec{Q})\hat{\rho }(-\vec{Q}) \right.}+ & \nonumber \\
& +{{W}_{a-a}}(\vec{Q})\left[ {{\cos }^{2}}(2{{v}_{p}}|x(0)|)\hat{D}(\vec{Q})\hat{D}(-\vec{Q}) \right.+ & \nonumber \\
& +{{\sin }^{2}}(2{{v}_{p}}|x(0)|)N\theta (\vec{Q})\theta (-\vec{Q})- & \nonumber \\
& -\cos (2{{v}_{p}}|x(0)|)\sin (2{{v}_{p}}|x(0)|)\times  &\\
& \times \sqrt{N}\left. (\hat{D}(\vec{Q})\theta (-\vec{Q})+\theta (\vec{Q})\hat{D}(-\vec{Q})) \right]+ & \nonumber \\
& +{{W}_{0-a}}(\vec{Q})\left[ \cos (2{{v}_{p}}|x(0)|) \right.\times  & \nonumber \\
& \times (\hat{\rho }(\vec{Q})\hat{D}(-\vec{Q})+\hat{D}(\vec{Q})\hat{\rho }(-\vec{Q}))- & \nonumber \\
& -\sin (2{{v}_{p}}|x(0)|)\sqrt{N}(\hat{\rho }(\vec{Q})\theta (-\vec{Q})+ & \nonumber \\
& +\theta (\vec{Q})\hat{\rho }(-\vec{Q}))]\} & \nonumber
\end{eqnarray}
The third transformation concerns the captured photons
\begin{eqnarray}
& {{{\hat{\tilde{H}}}}_{0,ph}}= & \nonumber \\
& ={{D}_{ph}}(\sqrt{{{N}_{p}}}|y(0)|){{{\hat{H}}}_{0,ph}}D_{ph}^{-1}(\sqrt{{{N}_{p}}}|y(0)|)= & \nonumber \\
& =\left( \hbar {{\omega }_{\frac{\pi }{{{L}_{c}}}}}-{{\mu }_{ph}} \right){{N}_{p}}|y(0){{|}^{2}}+ & \nonumber \\
& +\sum\limits_{{{{\vec{k}}}_{||}}}{\left( \hbar {{\omega }_{{\vec{k}}}}-{{\mu }_{ph}} \right)}c_{\vec{k},-}^{\dagger }c_{\vec{k},-}^{{}}- & \nonumber \\
& -\sqrt{{{N}_{p}}}|y(0)|\left( \hbar {{\omega }_{\frac{\pi }{{{L}_{c}}}}}-{{\mu }_{ph}} \right)\left( {{e}^{-i\beta }}c_{\frac{\pi }{{{L}_{c}}},-}^{\dagger }+{{e}^{i\beta }}c_{\frac{\pi }{{{L}_{c}}},-}^{{}} \right), & \nonumber \\
& \vec{k}=\frac{\pi }{{{L}_{c}}}{{{\vec{a}}}_{3}}+{{{\vec{k}}}_{||}}, & \\
& {{{\vec{k}}}_{||}}={{{\vec{a}}}_{1}}{{k}_{x}}+{{{\vec{a}}}_{2}}{{k}_{y}} & \nonumber
\end{eqnarray}
The last transformation involves the magnetoexciton and photons operators as follows
\begin{widetext}
\begin{eqnarray}
& {{{\hat{\tilde{H}}}}_{mex-ph}}=D(\sqrt{{{N}_{p}}}){{{\hat{H}}}_{mex-ph}}{{D}^{-1}}(\sqrt{{{N}_{p}}})= & \nonumber\\
& {{D}_{ex}}(\sqrt{{{N}_{p}}}|x(0)|)\{{{{\hat{H}}}_{mex-ph}}-\sqrt{{{N}_{p}}}|y(0)|[\varphi (0){{e}^{-i\beta }}(\vec{\sigma }_{\frac{\pi }{{{L}_{c}}}}^{+}\bullet \vec{\sigma }_{{{M}_{h}}}^{*})\hat{\Psi }_{ex}^{\dagger }(0)+ & \nonumber\\
& +{{\varphi }^{*}}(0){{e}^{i\beta }}(\vec{\sigma }_{\frac{\pi }{{{L}_{c}}}}^{-}\bullet \vec{\sigma }_{{{M}_{h}}}^{{}})\hat{\Psi }_{ex}^{{}}(0)]\}D_{ex}^{-1}(\sqrt{{{N}_{p}}}|x(0)|)= & \nonumber\\
& =-\sqrt{{{N}_{p}}}|y(0)|\{\hat{\Psi }_{ex}^{\dagger }(0)[\frac{1}{2}(\cos (2{{v}_{p}}|x(0)|)+1)\varphi (0){{e}^{-i\beta }}(\vec{\sigma }_{\frac{\pi }{{{L}_{c}}}}^{+}\bullet \vec{\sigma }_{{{M}_{h}}}^{*})+ & \nonumber\\
& +\frac{1}{2}(\cos (2{{v}_{p}}|x(0)|)-1){{\varphi }^{*}}(0){{e}^{i\beta -2i\alpha }}(\vec{\sigma }_{\frac{\pi }{{{L}_{c}}}}^{-}\bullet \vec{\sigma }_{{{M}_{h}}}^{{}})]+ & \nonumber\\
& +\hat{\Psi }_{ex}^{{}}(0)[\frac{1}{2}(\cos (2{{v}_{p}}|x(0)|)+1){{\varphi }^{*}}(0){{e}^{i\beta }}(\vec{\sigma }_{\frac{\pi }{{{L}_{c}}}}^{-}\bullet \vec{\sigma }_{{{M}_{h}}}^{{}})+ &\\
& +\frac{1}{2}(\cos (2{{v}_{p}}|x(0)|)-1)\varphi (0){{e}^{-i\beta +2i\alpha }}(\vec{\sigma }_{\frac{\pi }{{{L}_{c}}}}^{+}\bullet \vec{\sigma }_{{{M}_{h}}}^{*})]+ & \nonumber \\
& +\frac{\hat{D}(0)}{2\sqrt{N}}\sin (2{{v}_{p}}|x(0)|)[\varphi (0){{e}^{-i\beta +i\alpha }}(\vec{\sigma }_{\frac{\pi }{{{L}_{c}}}}^{+}\bullet \vec{\sigma }_{{{M}_{h}}}^{*})+{{\varphi }^{*}}(0){{e}^{i\beta -i\alpha }}(\vec{\sigma }_{\frac{\pi }{{{L}_{c}}}}^{-}\bullet \vec{\sigma }_{{{M}_{h}}}^{{}})]\}+ & \nonumber \\
& +\frac{1}{2}(\cos (2{{v}_{p}}|x(0)|)+1)\sum\limits_{{{{\vec{k}}}_{||}}}{[\varphi ({{{\vec{k}}}_{||}})(\vec{\sigma }_{{\vec{k}}}^{+}\bullet \vec{\sigma }_{{{M}_{h}}}^{*}){{c}_{\vec{k},-}}\Psi _{ex}^{\dagger }({{{\vec{k}}}_{||}})+{{\varphi }^{*}}({{{\vec{k}}}_{||}})(\vec{\sigma }_{{\vec{k}}}^{-}\bullet \vec{\sigma }_{{{M}_{h}}}^{{}})c_{\vec{k},-}^{\dagger }\Psi _{ex}^{{}}({{{\vec{k}}}_{||}})]}+ & \nonumber \\
& +\frac{1}{2}(\cos (2{{v}_{p}}|x(0)|)-1)\sum\limits_{{{{\vec{k}}}_{||}}}{{}}[\varphi ({{{\vec{k}}}_{||}}){{e}^{2i\alpha }}(\vec{\sigma }_{{\vec{k}}}^{+}\bullet \vec{\sigma }_{{{M}_{h}}}^{*}){{c}_{\vec{k},-}}\Psi _{ex}^{{}}(-{{{\vec{k}}}_{||}})+ & \nonumber \\
& +{{\varphi }^{*}}({{{\vec{k}}}_{||}}){{e}^{-2i\alpha }}(\vec{\sigma }_{{\vec{k}}}^{-}\bullet \vec{\sigma }_{{{M}_{h}}}^{{}})c_{\vec{k},-}^{\dagger }\Psi _{ex}^{\dagger }(-{{{\vec{k}}}_{||}})]+ & \nonumber \\
& +\frac{1}{2}\sin (2{{v}_{p}}|x(0)|)\sum\limits_{{{{\vec{k}}}_{||}}}{[\varphi ({{{\vec{k}}}_{||}}){{e}^{i\alpha }}(\vec{\sigma }_{{\vec{k}}}^{+}\bullet \vec{\sigma }_{{{M}_{h}}}^{*}){{c}_{\vec{k},-}}\frac{\hat{D}({{{\vec{k}}}_{||}})}{\sqrt{N}}+{{\varphi }^{*}}({{{\vec{k}}}_{||}}){{e}^{-i\alpha }}(\vec{\sigma }_{{\vec{k}}}^{-}\bullet \vec{\sigma }_{{{M}_{h}}}^{{}})c_{\vec{k},-}^{\dagger }\frac{\hat{D}_{{}}^{\dagger }({{{\vec{k}}}_{||}})}{\sqrt{N}}]} & \nonumber
\end{eqnarray}
\end{widetext}
Taking into account the relation (102) between the phases $\alpha$ ,$\beta$ and $\gamma $ as well the definition (110) of the operator $\theta (0)$ one can represent the transformed Hamiltonian with the broken gauge symmetry in the form
\begin{widetext}
\begin{eqnarray}
& \hat{\tilde{H}}={{N}_{p}}{{\left| y(0) \right|}^{2}}\left( \hbar {{\omega }_{\frac{\pi }{{{L}_{c}}}}}-{{\mu }_{ph}} \right)-\sqrt{{{N}_{p}}}\left| y(0) \right|\left( \hbar {{\omega }_{\frac{\pi }{{{L}_{c}}}}}-{{\mu }_{ph}} \right)\left( {{e}^{-i\beta }}c_{\frac{\pi }{{{L}_{c}}},-}^{\dagger }+{{e}^{i\beta }}c_{\frac{\pi }{{{L}_{c}}},-}^{{}} \right)- & \nonumber\\
& -\hat{\theta }(0)\sqrt{{{N}_{p}}}\left[ \frac{1}{2}\left( {{E}_{mex}}({{S}_{e}},{{R}_{1}};{{S}_{h}},{{M}_{h}},\varepsilon _{m}^{-})-{{\mu }_{ex}} \right) \right.\sin (2{{v}_{p}}\left| x(0) \right|)+ & \nonumber\\
& \left. +{{v}_{p}}\left| y(0) \right|\left| \varphi (0)\left( \vec{\sigma }_{\frac{\pi }{{{L}_{c}}}}^{-}\bullet \vec{\sigma }_{{{M}_{h}}}^{{}} \right) \right|\cos (2{{v}_{p}}\left| x(0) \right|) \right]+ & \nonumber\\
& +\frac{\hat{D}(0)}{2}\left[ {{E}_{mex}}({{S}_{e}},{{R}_{1}};{{S}_{h}},{{M}_{h}},\varepsilon _{m}^{-}) \right.-{{\mu }_{ex}}-2{{v}_{p}}\sin (2{{v}_{p}}\left| x(0) \right|)\left| y(0) \right|\left| \varphi (0)\left( \vec{\sigma }_{\frac{\pi }{{{L}_{c}}}}^{-}\bullet \vec{\sigma }_{{{M}_{h}}}^{{}} \right) \right|]+ & \nonumber\\
& +\left( {{G}_{e-h}}({{S}_{e}},{{R}_{1}};{{S}_{h}},{{M}_{h}},\varepsilon _{m}^{-})-{{\mu }_{e}}+{{\mu }_{h}} \right)\frac{\hat{\rho }(0)}{2}+\sum\limits_{{{{\vec{k}}}_{||}}}{\left( \hbar {{\omega }_{{\vec{k}}}}-{{\mu }_{ph}} \right)}c_{\vec{k},-}^{\dagger }c_{\vec{k},-}^{{}}+ & \nonumber\\
& +\frac{1}{2}\sum\limits_{{\vec{Q}}}{\left\{ {{W}_{0-0}}(\vec{Q})\hat{\rho }(\vec{Q})\hat{\rho }(-\vec{Q}) \right.}+{{W}_{a-a}}(\vec{Q})\left[ {{\cos }^{2}}(2{{v}_{p}}|x(0)|)\hat{D}(\vec{Q})\hat{D}(-\vec{Q}) \right.+ & \nonumber\\
& +{{\sin }^{2}}(2{{v}_{p}}|x(0)|)N\hat{\theta }(\vec{Q})\hat{\theta }(-\vec{Q})- &\\
& -\frac{1}{2}\sin (4{{v}_{p}}|x(0)|)\sqrt{N}\left. (\hat{D}(\vec{Q})\hat{\theta }(-\vec{Q})+\hat{\theta }(\vec{Q})\hat{D}(-\vec{Q})) \right]+ & \nonumber\\
& +{{W}_{0-a}}(\vec{Q})\left[ \cos (2{{v}_{p}}|x(0)|) \right.(\hat{\rho }(\vec{Q})\hat{D}(-\vec{Q})+\hat{D}(\vec{Q})\hat{\rho }(-\vec{Q}))- & \nonumber\\
& \left. -\left. \sin (2{{v}_{p}}|x(0)|)\sqrt{N}(\hat{\rho }(\vec{Q})\hat{\theta }(-\vec{Q})+\hat{\theta }(\vec{Q})\hat{\rho }(-\vec{Q})) \right] \right\}+ & \nonumber\\
& +\frac{1}{2}(\cos (2{{v}_{p}}|x(0)|)+1)\sum\limits_{{{{\vec{k}}}_{||}}}{[\varphi ({{{\vec{k}}}_{||}})(\vec{\sigma }_{{\vec{k}}}^{+}\bullet \vec{\sigma }_{{{M}_{h}}}^{*}){{c}_{\vec{k},-}}\Psi _{ex}^{\dagger }({{{\vec{k}}}_{||}})+{{\varphi }^{*}}({{{\vec{k}}}_{||}})(\vec{\sigma }_{{\vec{k}}}^{-}\bullet \vec{\sigma }_{{{M}_{h}}}^{{}})c_{\vec{k},-}^{\dagger }\Psi _{ex}^{{}}({{{\vec{k}}}_{||}})]}+ & \nonumber\\
& +\frac{1}{2}(\cos (2{{v}_{p}}|x(0)|)-1)\sum\limits_{{{{\vec{k}}}_{||}}}{{}}[\varphi ({{{\vec{k}}}_{||}}){{e}^{2i\alpha }}(\vec{\sigma }_{{\vec{k}}}^{+}\bullet \vec{\sigma }_{{{M}_{h}}}^{*}){{c}_{\vec{k},-}}\Psi _{ex}^{{}}(-{{{\vec{k}}}_{||}})+ & \nonumber\\
& +{{\varphi }^{*}}({{{\vec{k}}}_{||}}){{e}^{-2i\alpha }}(\vec{\sigma }_{{\vec{k}}}^{-}\bullet \vec{\sigma }_{{{M}_{h}}}^{{}})c_{\vec{k},-}^{\dagger }\Psi _{ex}^{\dagger }(-{{{\vec{k}}}_{||}})]+ & \nonumber\\
& +\frac{1}{2}\sin (2{{v}_{p}}|x(0)|)\sum\limits_{{{{\vec{k}}}_{||}}}{[\varphi ({{{\vec{k}}}_{||}}){{e}^{i\alpha }}(\vec{\sigma }_{{\vec{k}}}^{+}\bullet \vec{\sigma }_{{{M}_{h}}}^{*}){{c}_{\vec{k},-}}\frac{\hat{D}({{{\vec{k}}}_{||}})}{\sqrt{N}}+{{\varphi }^{*}}({{{\vec{k}}}_{||}}){{e}^{-i\alpha }}(\vec{\sigma }_{{\vec{k}}}^{-}\bullet \vec{\sigma }_{{{M}_{h}}}^{{}})c_{\vec{k},-}^{\dagger }\frac{\hat{D}_{{}}^{\dagger }({{{\vec{k}}}_{||}})}{\sqrt{N}}]} & \nonumber
\end{eqnarray}
\end{widetext}
Looking at this expression one may conclude that side by side with the u-v-type transformation (110) of the magnetoexciton superposition-type operator $\hat{\theta }(\vec{Q})$ and of the acoustical plasmon density operator $\hat{D}(\vec{Q})/\sqrt{N}$ another mixed state of the type acoustical plasmon-polariton under the influence of the magnetoexciton-polariton BEC appeared. In addition to them there are anti-resonance-type terms in the magnetoexciton-photon interaction, even if they were not included in the initial Hamiltonian (87). The obtained results permit to determine the chemical potentials ${{\mu }_{ex}}$ and ${{\mu }_{ph}}$ and to investigate the energy spectrum of the collective elementary excitations.
\section{Conclusions:}
The influence of the RSOC on the properties of the 2D magnetoexcitons was described taking into account the results concerning the Landau quantization of the 2D electrons and holes with nonparabolic dispersion laws, pseudospin components and chirality terms [18, 19, 22]. The main attention was paied to the study of the operators $\hat{\rho }(\vec{Q})$ and $\hat{D}(\vec{Q})$, which together with the magnetoexciton creation and annihilation operators $\hat{\Psi }_{ex}^{\dagger }({{\vec{k}}_{||}})$ and $\hat{\Psi }_{ex}^{{}}({{\vec{k}}_{||}})$ form a set of four two-particle integral operators. It was shown that the Hamiltonians of the electron-radiation and Coulomb electron-electron interactions can be expressed in the terms of these four integral two-particles operators. The unitary transformation breaking the gauge symmetry of the deduced Hamiltonian and the BEC of the magnetoexciton-polaritons were introduced in the frame of the Keldysh-Kozlov-Kopaev method using the polariton creation and annihilation operators. They were expressed through the same magnetoexciton and photon operators using the Hopfield coefficients in a simplified form without the anti-resonance terms because the energies of the participant quasiparticles are finite situated near the energy of the cavity mode. The unitary transformation is factorized as a product of two unitary transformations acting independently in two magnetoexciton and photon subsystems. It was realized that the BEC of magnetoexciton polaritons supplementary gives rise to the acoustical plasmon-photon interaction and to a new type plasmon-polariton formation. The antiresonance terms of the magnetoexciton-photon interaction also appeared even if they were neglected in the starting Hamiltonian. The mixed magnetoexciton-acoustical plasmon states in the absence of the RSOC were investigated in Refs[32-34]. The obtained final transformed Hamiltonian will be used to study the collective elementary excitations.
\acknowledgments{E.V.D., I.V.P. and V.M.B thanks the Foundation for Young Scientists of the Academy of Sciences of Moldova for financial support (14.819.02.18F). S.S.R and E.V.D. thanks the Academy of Sciences of Moldova and Federal Ministry of Education and Research of Germany(BMBF) for financial support(13.820.05.08/GF)}

\end{document}